\documentclass{emulateapj}
\usepackage{amsmath}

\usepackage{color} 
\slugcomment{For Submission to ApJ}
\shorttitle{The Blazar Envelope}
\shortauthors{Meyer et al.}

\begin{document}

\title{From the Blazar Sequence to the Blazar Envelope: \\
    Revisiting the Relativistic Jet Dichotomy in Radio-loud AGN}

\author{Eileen T. Meyer and Giovanni Fossati}%\altaffilmark{1}}
\email{meyer@rice.edu}
\affil{Department of Physics and Astronomy, Rice University,
    Houston, TX 77005}

\author{Markos Georganopoulos}%\altaffilmark{2}}
\affil{Department of Physics, University of Maryland Baltimore County, Baltimore, MD 21250}
\affil{NASA Goddard Space Flight Center, Mail Code 663, Greenbelt, MD 20771, USA}
\author{Matthew L. Lister}%\altaffilmark{3}}
\affil{Department of Physics, Purdue University, West Lafayette, IN 47907}

%\altaffiltext{1}{Rice University Houston, TX 77005}
%\altaffiltext{2}{UMBC Baltimore, MD 21250}
%\altaffiltext{3}{Purdue University, West Lafayette, IN 47907}

\begin{abstract}
We revisit the concept of a blazar sequence that relates the
synchrotron peak frequency ($\nu_{\mathrm{peak}}$) in blazars with
synchrotron peak luminosity ($\mathrm{L}_{\mathrm{peak}}$, in
$\nu\mathrm{L}_\nu$) using a large sample of radio-loud AGN. We
present observational evidence that the blazar sequence is formed from
two populations in the synchrotron $\nu_{\mathrm{peak}}$ $-$
$\mathrm{L}_{\mathrm{peak}}$ plane, each forming an upper edge to an
envelope of progressively misaligned blazars, and connecting to an
adjacent group of radio galaxies having jets viewed at much larger
angles to the line of sight. When binned by jet kinetic power
($\mathrm{L}_\mathrm{kin}$; as measured through a scaling relationship
with extended radio power), we find that radio core dominance
decreases with decreasing synchrotron $\mathrm{L}_{\mathrm{peak}}$,
revealing that sources in the envelope are generally more
misaligned. We find population-based evidence of velocity gradients in
jets at low kinetic powers ($\sim 10^{42}-10^{44.5}$ erg s$^{-1}$),
corresponding to FR I radio galaxies and most BL Lacs.  These low jet
power `weak jet' sources, thought to exhibit radiatively inefficient
accretion, are distinguished from the population of non-decelerating,
low synchrotron-peaking (LSP) blazars and FR II radio galaxies
(`strong' jets) which are thought to exhibit radiatively {\it
  efficient} accretion. The two-population interpretation explains the
apparent contradiction of the existence of highly core-dominated,
low-power blazars at both low and high synchrotron peak frequencies,
and further implies that most intermediate synchrotron peak (ISP)
sources are not intermediate in intrinsic jet power between LSP and
high synchrotron-peaking (HSP) sources, but are more misaligned
versions of HSP sources with similar jet powers.
\end{abstract}


\section{INTRODUCTION}
\label{intro}

The success of the current model for radio-loud Active Galactic Nuclei
(AGN) as accreting super-massive black holes with bipolar,
relativistic jets has rested on the ability of this physical paradigm
to successfully explain the chameleon nature of the spectral energy
distribution (SED) of this family of extreme objects. The jet model is
inherently anisotropic and allows us to unify the highly luminous and
variable blazars as relativistically beamed counterparts of
Fanaroff-Riley (FR) radio galaxies \citep{fan74}.

The typical blazar spectrum exhibits a double peak (in $\nu F_{\nu}$),
attributed to synchrotron and inverse Compton (IC) radiation from the
relativistic jet, which dominates in blazars due to the alignment of
the jet axis near to our line-of-sight \cite[for a review of leptonic
blazar models, see][]{bot07}. The synchrotron component peaks from
sub-infrared (IR) energies up to hard X-rays, with the IC component
peaking at much higher energies (typically
gamma-rays). Spectroscopically, blazars are traditionally divided
between optically `featureless' BL\,Lac objects and (broad) emission
line-emitting Flat Spectrum Radio Quasars (FSRQ). The long-standing
orientation-based unification scheme for radio-loud AGN proposes BL
Lacs and FSRQ as counterparts to FR I and FR II radio galaxies,
respectively, based on similar spectra, morphologies, and range in
extended radio luminosity \citep{urr95}.

However, violations to this scheme are well-known, and findings of
powerful, FR II-like BL Lacs, and low-power FSRQ
\citep{lan06,lan08,kha10} appear to break the simplest dichotomy
(going back to the original FR classification) between the
lobe-dominated, high-power FR II galaxies and the low-power,
edge-darkened FR Is. \cite{kha10} also found BL Lacs displaying SEDs
and hot-spots typical of quasars while others have observed BL Lacs
exhibiting broad lines in low continuum states
\citep{ver95,cor00,rai07}.

Further, the importance of optical spectral type in connecting radio
galaxies with their aligned counterparts is difficult to assess due to
the strong effects of relativistic beaming on equivalent width
measurements on which the blazar classifications are based
\citep{geo98_thesis,ghi11_bllacs_fsrq}.  At the same time, the
designation of Low and High Excitation Radio Galaxies (LERG and HERG,
corresponding to prominence of emission lines as measured with an
excitation index, see \citealt{hin79}) is increasingly found to be
mixed among FR I and FR II morphological types; all of this creates
serious problems for a unification scheme which connects spectral type
with the morphology.

\subsection{The Blazar Sequence}

The synchrotron peak frequency ($\nu_{\mathrm{peak}}$) may take on a wide range of
values (from 10$^{12}$ to 10$^{18}$ Hz at the extremes), and is one of
the principle ways to classify individual blazars. While FSRQ sources
are typically found to have low $\nu_{\mathrm{peak}}$ ($<$ 10$^{14.5}$
Hz), BL Lacs are found to span the entire range, with low,
intermediate, and high-peaking BL Lacs (LBL, IBL, and HBL) typically
dominating surveys selecting in the radio, optical, or X-ray,
respectively. We adopt the generic terms for low, intermediate, and
high {\sl synchrotron-peaking} (LSP, ISP, HSP) blazars independently
of the spectroscopic type, as in \cite{abd10_LSP_ref}.

\cite{fos98} found an anti-correlation between the frequency of the
synchrotron peak and the bolometric and synchrotron peak luminosities
and suggested a link between the power of the source and
$\nu_{\mathrm{peak}}$. This sequence appeared to match a spectroscopic
one, from predominantly powerful FSRQ sources at the low-peak,
high-luminosity end through LBLs and finally HBLs, at the
low-luminosity end.  \cite{ghi98} suggested that more efficient
cooling of particles in the jets of high luminosity blazars is
responsible for the lower peak frequencies. However, it is not clear
how the continuous blazar sequence fits in with the bi-modality of
morphology and spectral type in radio galaxies.

\subsection{Beaming and Quantitative Unification}
\label{velocitygrad}

Orientation-based unification predicts that misaligned blazars will
drop in luminosity and synchrotron peak frequency according to the
decrease in Doppler boosting with increasing angle.  In the simplest
case (a single Lorentz factor and a convex spectrum), the ratio of
decrease in log frequency to log luminosity will be approximately 1:4,
roughly orthogonal to the sequence. \citep[This holds for moving
features in the jet; for a standing shock model the relation is 1:3,
see {\sl e.g.,}][]{lin85}. The connection between blazars and the
{\sl nuclear emission} in radio galaxies can naturally inform our jet
models.  The synchrotron nature of the nuclear radio galaxy emission
has been confirmed by comparison of the broad-band spectral indices
and SED characteristics of FR I nuclei with BL Lacs
\citep{chi99,cap00,chi00,cap02,tru03}, where the effects of
relativistic beaming (rather than obscuration) accounted for the
10$-$10$^4$ difference in radio and optical luminosities between the
BL Lacs and radio galaxies. 

\cite{cap02} and \cite{tru03} agree upon jet bulk Lorentz factors of
$\Gamma$ $\sim$ 4$-$6 in order to account for the difference in radio
galaxy and blazar luminosities, while \cite{har03} found $\Gamma$
$\sim$ 3 sufficient to match the luminosity functions of the B2 and
EMSS and 1 Jy samples. These values agree with VLBI studies of the
pc-scale jets of HSP BL Lacs which show that they are barely
super-luminal, with average bulk Lorentz factors $\Gamma$ $\sim$ 2$-$3
\citep{pin01_tev_bllacs,edw02_tev_bllacs,gir04,pin08,pin10}; however,
several LSP BL Lacs show high velocities on VLBI scales
\citep{jor01,coh07}. Interestingly, \cite{har03} also found that they
could not reproduce the redshift distribution of 1 Jy BL Lac sample
unless some of the high-redshift BL Lacs corresponded to FR IIs.

In contrast, high Doppler factors ($\sim$ 50 or greater) must be
invoked to explain the brightness and rapid variability of TeV
blazars, as the jet must be highly relativistic in order to avoid
absorption by co-spatial IR photons
\citep{don95,aha07,ghi09_tev_variability,gia09}. The discrepancy in
apparent jet speeds and radio galaxy nuclear emission can be resolved
at least for FR I/BL Lac sources by invoking the presence of
structured jets with multiple emitting regions of varying speeds.
This includes the spine-sheath axisymmetric model explored by
\cite{swa98}, \cite{chi00}, and \cite{ghi05}, and the decelerating
model of \cite{geo03}, all of which imply a high-speed spine or base
which dominates the flux when the source is aligned, and a slower
component which dominates if the source is misaligned. Other blazar
studies also support the decelerating model for low power blazars, as
recent VLBI observations of the MOJAVE sample\footnote{MOJAVE
  (Monitoring Of Jets in Active galactic nuclei with VLBA
  Experiments): Over 250 bright blazars being continuously monitored
  with VLBI.  (www.physics.purdue.edu/mojave)} reveal that the these
sources also have lower apparent jet speeds, versus the high-power
FSRQ which maintain collimated jets and consistent jet speeds over
longer distances \citep{lis09}.

\subsection{The Blazar Envelope}
The simplest scenario behind the appearance of a blazar sequence is
that the physical properties of extragalactic jets (as measured
through the synchrotron peak frequency and luminosity) are a function
of a single parameter, the jet kinetic power
($\mathrm{L}_\mathrm{kin}$). If there is a one-to-one correspondence
between $\nu_\mathrm{peak}$ and $\mathrm{L}_\mathrm{kin}$, the
original blazar sequence could be seen as the extreme `edge' of an
envelope, consisting of the most aligned (and therefore brightest)
objects.  More misaligned sources should appear below the
sequence, causing it to widen.  If the jets are monoparametric in
terms of jet power, these misaligned sources should fall in a line
away from their aligned counterparts on the sequence, according to the
drop in Doppler factor, forming `tracks' of similar kinetic power in
the envelope of blazars populating the $\nu_\mathrm{peak}$ -
$\mathrm{L}_\mathrm{peak}$ plane.

The original blazar sequence was based on 126 objects, a few dozen of
which were detected in gamma-rays. Over a decade later there are
thousands more blazars and blazar candidates in the literature and a
much greater density of multi-wavelength data in general. Many of the
newer samples have been designed to combat the early selection effects
of the radio and X-ray surveys, such as the low radio power {\it
  CLASS} sample \citep{cac04_class_testing_the_sequence}, and it is
now clear that as our surveys go deeper, many more low-luminosity
blazars are emerging. These newer samples challenge both the sequence
and its simplest extension that considers the sources in the envelope
simply to be de-beamed blazars. Contrary to what is anticipated by the
sequence, \cite{lan08} showed that HSP and LSP BL Lacs objects cover
similar ranges in extended radio power.
\cite{cac04_class_testing_the_sequence} found several low luminosity,
low $\nu_\mathrm{peak}$ sources with relatively high core dominance
compared to brighter objects at similar low $\nu_\mathrm{peak}$,
suggesting that the former are not simply related to the latter
through misalignment, as one might naively expect.

In this paper we examine the available evidence for a `blazar
envelope' beneath the upper-limit of the most aligned blazars, due
primarily to mis-alignment. We test the possibility that jets are
monoparametric engines by examining how the source power is related to
other properties of the blazar such as location in the
$\nu_{\mathrm{peak}}-\mathrm{L}_{\mathrm{peak}}$ plane, and
re-evaluate the blazar sequence in light of the great increases in
multi-wavelength, multi-epoch data available for blazar and radio galaxy
jets.

% -------------------------------------------------------------------------------
\begin{deluxetable*}{l l c p{7cm}}
%\rotate
%\begin{deluxetable}{l l c l}
  \tablecaption{Blazar Samples\label{table:sources}}
  \tablewidth{0pt}
  \tablehead{
    \colhead{Sample} & \colhead{Abbrev.} & \colhead{No. of}  & \colhead{References}\\
    & & Objects\tablenotemark{$\dagger$} & \\
  }
  \startdata
  1 Jansky blazar sample & 1 Jy & 34 & \cite{sti91_1jybll} \\
  2-degree field (2dF) QSO survey & 2QZ & 56 & \cite{lon02_2qz_i,lon07_2qz} \\
  2 Jansky survey of flat-spectrum sources& 2 Jy & 213 &  \cite{wal85} \\
  The Candidate Gamma-Ray Blazar Survey & CGRaBS & 1625 &  \cite{hea08_cgrabs} \\
  Cosmic Lens All-sky Survey & CLASS & 236 & \cite{mar01_class_i,cac02_class_ii,cac04_class_testing_the_sequence} \\
  Deep X-ray Radio Blazar Survey & DXRBS & 283 & \cite{per98_dxrbs_i,lan01_dxrbs_ii,pad07_dxrbs_iii}\\
  \emph{EGRET} blazar sample & ... & 555 &  \cite{muk97_egret_blazars_update} \\
  \emph{Einstein} Slew survey sample of BL Lac objects & ... &48 &  \cite{per96_slewbll} \\
  \emph{Fermi} blazar sample & 1FGL & 689 &  \cite{abd10_fermi_cat} \\
  Hamburg-RASS bright X-ray AGN sample & HRX & 104 & \cite{bec03_hrx_evolution} \\
  Parkes quarter-Jansky flat-spectrum sample & ... & 878 & \cite{jac02_parkes_i, hoo03_parkes_ii, wal05_parkes_iii} \\
  Radio-Emitting X-ray source survey & REX & 55 &  \cite{cac99_rex_i} \\
  Radio-Optical-X-ray catalog  & ROXA & 816 &  \cite{tur07_roxa} \\
  RASS - Green Bank BL Lac sample & RGB & 127 & \cite{lau99_rgb_sample} \\
  \emph{Einstein} Medium-Sensitivity Survey of BL Lacs & EMSS & 44 & \cite{mor91_emss_bllacs,rec00_emssbll_ii} \\
  RASS - SDSS flat-spectrum sample & ... & 501 & \cite{plo08_sf}\\
  Sedentary survey of high-peak BL Lacs & ... & 150 &  \cite{gio99_sed_i,gio05_sed_ii, pir07_sed_iii} \\ 
  \enddata
  \tablenotetext{$\dagger$}{Number of sources listed is total sources in sample (not necessarily number of blazars).}
\end{deluxetable*}
% -----------------

The paper is organized as follows: in \S \ref{methods} we explain the
sample selection and describe our statistical and phenomenological
models which allowed us to estimate the important envelope parameters
for blazars and radio galaxies. In \S \ref{results} we present the
results of the first round of data collection and analysis and the
`blazar envelope'. In \S \ref{discussion} we discuss the implications
of the blazar envelope and the potential for
constraining jet models. In \S \ref{conclusions}, we summarize our
findings. Throughout the paper we adopt a cosmology based on the
concordance model with $\Omega_M$ = 0.27, $\Omega_{\Lambda}$ = 0.73,
and $H_0 = 71 \rm ~ km ~ s^{-1} ~ Mpc^{-1}$. Energy spectral indices
are defined such that $f_\nu \propto \nu^{-\alpha}$, and all
luminosities (L$_\nu$) are references to powers ($\nu$L$_\nu$
implied).

\section{METHODS}
\label{methods}

In order to investigate the nature of the blazar envelope, we require
a sample of blazars with a well-sampled SED such that a reliable
estimate of the synchrotron peak location
%($\nu_{\mathrm{peak}}$ and $\mathrm{L}_{\mathrm{peak}}$) 
can be made. The scarcity of such blazars may recommend utilizing as
many sources as are known without regard for flux-limited samples, but
this approach risks losing the ability to understand the limits of our
observations.  We compromise by combining multiple well-defined
(flux-limited) samples of flat-spectrum sources. The overall sample
definition is thus complex, but does allow for the examination of
biases by population modeling, as discussed in \S
\ref{section:biases}.

\subsection{The Blazar Sample}
The flat spectrum candidate sample was made from the surveys listed in
Table \ref{table:sources}. A multi-frequency database of all sources
in this master list was compiled by a combination of all the data in
the original publications, with additional data from the literature
and from the SIMBAD and NED online databases, for a total of 3773
sources (taking into account duplicates). For 69 sources, previously
unpublished Swift data were available which were utilized to produce
fluxes and X-ray indices. Spectroscopic classification was taken from
NED, SIMBAD, or the most recent publication, and non-blazar sources
were removed, leaving 3014 blazars.  A further reduction was made by
removing sources with sparse multi-frequency SED sampling. The minimum
required sampling was photometry at 3, 2, and 1 or more unique
frequencies for the (rest frame) frequency ranges $< $ 10$^{11.5}$,
10$^{11.5}$$-$10$^{16}$, and $> $ 10$^{16}$ Hz, respectively, yielding
1327 sources suitable for further analysis. We discuss the properties
of the discarded sources in \S \ref{sec:discards}.

For sources with unknown redshift ($\sim$ 16\% of the sample), we use
the relationship between optical host galaxy magnitude and redshift of
\cite{sba05} to estimate a value for these sources, taking the lower
limit when exact host magnitudes were unavailable. All luminosities
and frequencies were corrected to rest-frame values before spectral
fitting; the global fits (rather than an assumed index) were then used
to find k-corrected rest-frame quantities of interest.

\subsection{Jet Power}
\label{sec:jetpower}
Measuring the jet kinetic jet power of our blazars and radio galaxies
allows us to see if the areas occupied by jets of similar power in the
$\nu_{\mathrm{peak}}-\mathrm{L}_{\mathrm{peak}}$ plane can be
understood simply as an orientation effect (as it would be anticipated
in the original blazar sequence), or if additional considerations are
needed.  We note that for our purposes we need only classify sources
into a few broad bins of jet power (as we show below, a width of one
decade gives us a sufficient number of well populated jet power bins),
and an extremely accurate estimation of the jet kinetic power is not
needed.
 
\cite{raw91_nature} estimated jet power by dividing the lobe energy
content of radio galaxies by their age. The energy content was
estimated through minimum energy arguments
\citep[\emph{e.g.}][]{mil80_radio_gal_review} derived from measurements of
the {\sl low frequency radio emission} where the lobe emission is
assumed to dominate over the jet, while the source age was estimated
from spectral breaks in the lobe spectra through synchrotron aging
arguments.  \cite{wil99_7c_lines} revisited the issue of obtaining the
jet power from the low frequency extended luminosity. They studied in
detail the parameters that come into this relation, including the
unknown energy content of the non-radiating particles and the time
evolution of the lobe luminosity \cite[see also][]{blundell99} and
presented a scaling relation which included a careful parametrization
of the various uncertainties (see their Section 4.1, and Equation 11).

The actual normalization for this scaling between the jet kinetic
power and the low frequency extended luminosity
($\mathrm{L}_\mathrm{ext}$) came from a very different route. The dark
cavities seen in (\emph{e.g.,} Chandra) X-ray images are created by
AGN jet activity and can be used to estimate the $p\Delta V$ work
needed to inflate them. Importantly, this energy estimate does not
depend on the minimum energy assumption and does not require knowledge
of the amount of non-radiating matter in the lobes. The jet power
$\mathrm{L}_{\mathrm{kin}}$ is estimated by dividing the $p\Delta V$
work needed to expand the cavity by the cavity age found by estimating
the lobe buoyancy time \citep{bir08,cav10}.

  While this method is limited to a small number of sources at
  present, the $\mathrm{L}_{\mathrm{kin}}$ $-$
  $\mathrm{L}_{\mathrm{ext}}$ relation covers almost six orders of
  magnitude in jet power, including both FR I and FR II sources. The
  slope of the scaling \cite{cav10} find is compatible within errors
  with that found by \cite{wil99_7c_lines} and requires a lobe energy
  content larger by at least a factor of 100 compared to that inferred
  by minimum energy arguments.  The cavity work shows convincingly
  that $\mathrm{L}_{\mathrm{kin}}$ and the extended radio luminosity
  are related. We therefore chose the low-frequency extended
  luminosity (also commonly referred to as steep or lobe emission) at
  300 MHz ($\mathrm{L}_{300}$) as an estimator of the jet power for
  this work. This value is then converted to
  $\mathrm{L}_{\mathrm{kin}}$ from the linear fit (with 1-$\sigma$
  error),
\begin{equation}
\begin{split}
  \label{eq:qjet}\mbox{log }\mathrm{L}_{\mathrm{kin}} = 0.64(\pm0.09)&\,\left(\mbox{log }\mathrm{L}_{300}-40\right)\\
  &+ 43.54(\pm0.12)\,\,\mbox{  [ergs s}^{-1}\mbox{]}\\
\end{split}
\end{equation}
taken from \cite{cav10}. Although the scatter in the scaling is not
insignificant ($\sim 0.7 $ decade in $\mathrm{L}_{\mathrm{kin}}$), it
is adequate for our purpose of classifying sources in broad bins of
jet power.  This value ($\mathrm{L}_{\mathrm{kin}}$) is what we
generally refer to throughout the paper as the jet (kinetic) power,
not to be confused with $\mathrm{L}_{\mathrm{peak}}$ or other
observables of the jet which are affected by beaming.

\subsubsection{Extended Radio Emission}
\label{section:VLA}

The extended radio emission from the lobes of radio galaxies differs
from the beamed core emission in being both non-variable and generally
exhibiting a much steeper spectrum. Thus one way to isolate the
extended emission is to look for a change in the spectral index at low
frequencies. For our sample, a spectral decomposition (SD) method was
applied to all sources which had data points at five or more
frequencies below $10^{11}$ Hz. Sources were tested for a
two-component spectrum by fitting alternately with a double power-law
and then a single power law (assuming normal error). Because the
latter can be expressed as a particular case of the former, this
allowed us to use the maximum likelihood (L) ratio test for nested
models. In this test, the statistic $-$2$\Delta$log(L) distribution
can be approximated as a $\chi^2$ with $n_2$ - $n_1$ degrees of
freedom \citep[][]{wil38}.  The double power-law model has $n_2$ = 4
degrees of freedom (slope and normalization for each powerlaw), while
the null hypothesis model has $n_1$ = 2, for a total of two degrees of
freedom. We rejected the two power-law fit for a significance level
$\alpha > $0.05. The best steep-spectrum fit from the hypothesis test
was then used to calculate $\mathrm{L}_{300}$.

The normal approximation for the spread around the power law is a
rough approximation, and for eight sources it appears the fit failed
to find a significant two-powerlaw solution due to a great deal of
variability in the radio core points. $\mathrm{L}_{300}$ was estimated
for these sources by fitting hand-selected low-frequency radio points
corresponding to a stable emission component persistent beneath the
core variability. This method was also used for a few sources which
passed the test, where core variability caused the steep power law fit
to be somewhat more flat than would be fit by eye (a correction of a
few percent).  We note that inclusion of these eight sources and the
corrections do not materially affect the results presented.

The SD method is the most reliable way to separate the boosted core
emission from the isotropic lobe emission, as it is model
independent. Another method is to measure the extended emission from
core-subtracted VLA images (typically at 1.4 GHz). In order to expand
our sample and counter some selection bias (see longer discussion in
\S \ref{section:biases}), extended flux measurements were taken from
the literature where available (see Table \ref{table:allsources} for
references). One difficulty with these measurements is that an
assumption must be made about the spectral index used to extrapolate
from 1.4 GHz to the 300 MHz luminosity. It is also clear from sources
with both a visible steep spectrum and VLA data that these are not
always in agreement. Even for the same object, VLA estimates by
different authors can vary widely, as in the case of 3C 446, with 1.4
GHz extended flux estimates of 91.6, 271, and 4030 mJy
\citep{kha10,coo07_mojave_iii,ant85}, none of which agrees with a
clear steep component which suggests a flux $\simeq$ 800 mJy at that
frequency. However, examination of the over 100 sources with both VLA
and SD estimates reveals that typical differences are less than 0.5 in
log luminosity, with the best agreement found using a low-frequency
index of $\alpha$ = 1.2 for the extrapolation to 300 MHz.

\begin{figure}
 \includegraphics[width=0.99\linewidth]{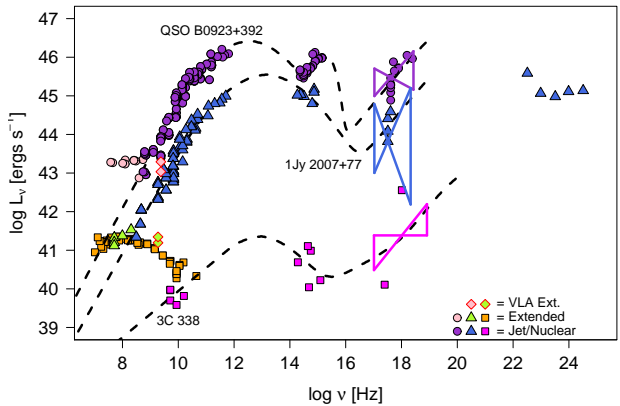}

  \caption{SEDs for three sources in which the extended radio emission
    is clearly visible as a steep component at low frequencies. The
    FSRQ QSO B0923+392 (top) has a very clear steep component at low
    frequencies, dominating up to $\sim$ 10$^{9.5}$ Hz, and a
    prominent `big blue bump' from the accretion disk, suggesting that
    it may be relatively misaligned compared with the
    \emph{Fermi}-detected BL Lac 1Jy 2007+77 (middle), because they
    differ by $\sim 2 $ decades in extended luminosity but only $\sim
    1$ decade in beamed power.  This latter source has a similar
    extended radio spectrum to the radio galaxy 3C 338 (bottom),
    however in the radio galaxy more of the extended emission is
    visible due to the nuclear emission being much less beamed. We
    also show estimates of the extended emission at 1.4 GHz as
    measured from VLA maps (diamonds), which generally agree with the
    spectral fits. The dashed curves are the best-fit of the blazar
    model described in the text and in Appendix \ref{app:SEDfit}.}
\label{fig:extvscore}
\end{figure}

The full SED for three sources is shown in Figure \ref{fig:extvscore}:
the bright quasar QSO B0923+392 (top), the LSP BL Lac 1Jy 2007+77
(middle), and the FR I radio galaxy 3C 338 (bottom).  In all cases,
extended radio emission was detected in our two-component SD test --
these points are shown in lighter colors of pink, green, and orange,
respectively. (Estimates of the extended luminosity derived from VLA
maps are also shown as diamonds).  We note that while 1Jy 2007+77 and
3C 338 have similar extended luminosities at 300 MHz, the much greater
core-dominance of the blazar is evident, and results in much less of
the extended emission being visible. Also, while the quasar is
evidently more powerful than the other two sources by two orders of
magnitude in $\mathrm{L}_{300}$, we see that it is likely more
misaligned than 1Jy 2007+77, from the greater visibility of the steep
component and the prominent accretion disk bump in the optical. This
relatively bright source was also not a significant detection in the
\emph{Fermi}/LAT 1-year catalog, unlike 1Jy 2007+77. This could be
explained if the gamma-ray emission is more beamed than that at lower
frequencies.

\subsubsection{Connection Between Extended and Unbeamed Core Emission}
\begin{figure}
  \includegraphics[width=0.99\linewidth]{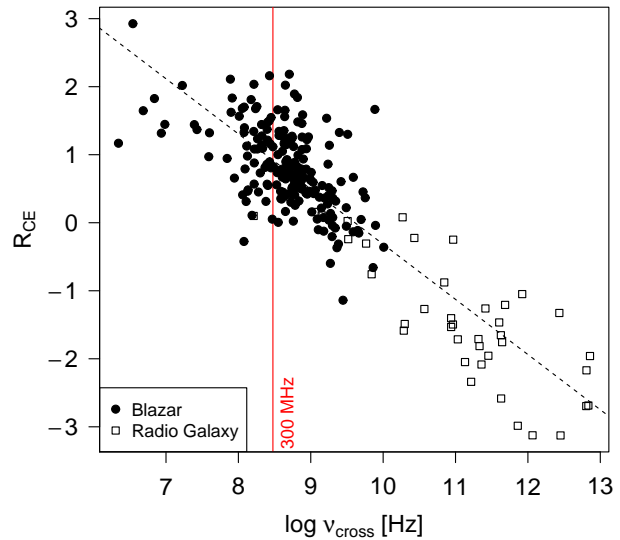}
  \caption{The crossing frequency, $\nu_{\mathrm{cross}}$ is the
    approximate frequency at which the SED becomes dominated by the
    beamed jet emission rather than the steep extended emission. Radio
    galaxies are shown as empty squares, and blazars as filled
    circles. As log core dominance ($\mathrm{R}_\mathrm{CE}$)
    increases due to beaming this crossing point moves to lower
    energies. Many sources will remain core-dominated to low
    frequencies, even to the limit of ground-based observations at
    $\sim$ 10 MHz. The broken line shown is the linear correlation
    between the plotted variables (r = 0.87). The solid red line marks
    the 300 MHz frequency at which the extended emission is measured.}
\label{fig:xcross}
\end{figure}

For the SD method, it is important to note that failure to find a two
power-law solution for a source can just as easily be due to a lack of
low-frequency data coverage as to the source emission being highly jet
dominated down to low frequencies. The approximate frequency at which
the beamed emission becomes dominant varies from $10^{6}$ to $10^{13}$
Hz in our sample. In Figure \ref{fig:xcross} we plot the `crossing'
frequency $\nu_{\mathrm{cross}}$ of the fitted jet SED and steep
component and the log core dominance $\mathrm{R}_{\mathrm{CE}}$ (we
define $\mathrm{R}_{\mathrm{CE}}$ as the log of the ratio of core to
extended luminosity at 1.4 GHz). The correlation is well-fit by a
linear regression model (correlation coefficient $r$ = 0.87). This
correlation can arise from two coexisting causes: an intrinsic spread
in unbeamed core power for a given extended luminosity and the effect
of beaming. If the effect of the natural spread of core power
dominates over the effects of beaming (as would be the case if beaming
of the core emission is weak or absent), then we expect to see radio
galaxies and blazars mixed along this correlation (this is because the
range by which each particular source moves before being classified as
a radio galaxy will be smaller than the initial intrinsic range that
the aligned blazars populate).  This is, however, not what we find, as
shown in Figure \ref{fig:xcross}. Instead, we observe a clear
separation of radio galaxies and blazars which suggests that the
effect of beaming dominates over the intrinsic spread and that the
range of unbeamed core emission for a given extended luminosity be
narrow.

\subsubsection{Caveats}

Many of our sources which lack VLA imaging have observations down to
very low frequencies (30 $-$ 200 MHz), and show no signs of a steep
component. It is impossible to tell if these sources are highly
core-dominated to the lowest observed frequencies or if the lobe
emission has a flat spectrum which is indistinguishable from the core
emission. The latter scenario may be a concern for the very low power
sources, as it is possible that the break frequency (\emph{i.e.}, the point
at which the lobe emission becomes steep) may increase with decreasing
synchrotron power as \cite{bru03} found for hot spots in radio
galaxies. For these sources it may be advisable to perform VLA map
subtractions at multiple frequencies to confirm such a spectrum. 

Finally, we must point out the risk in the common practice of assuming
that flux below $\sim$ 300 MHz, taken without examination of the
spectral characteristics, is actually from the extended isotropic
component, rather than the core. This assumption would be clearly
inappropriate for many of our sources, as shown by Figure
\ref{fig:xcross}, and could lead both to an overestimate of the jet
power and an underestimate of the radio core dominance often used to
infer the degree of mis-orientation of the source.

\subsection{Fitting the Synchrotron Spectrum}
For the 1327 blazars making the initial cut based on SED sampling, we
utilized a phenomenological parametric model in order to estimate the
synchrotron peak frequency and luminosity. The basic shape is
parabolic, with a power-law extension to low frequencies and a rising
power-law to simulate the beginning of the IC spectrum in the X-rays
\citep{fos97}. These components were free to vary within a reasonable
range in shape or slope, respectively, and in relative normalization,
for a total of 6 parameters. In sources where a big blue bump was
suggested on visual inspection, this feature was added to the SED fit
by adding a template accretion-disk spectrum \citep[see
][]{fra02_accretion_power_book}, adding one additional parameter for
the bump peak luminosity (see Appendix \ref{app:SEDfit} for a summary
of the model).

\begin{figure}
 \includegraphics[width=0.89\linewidth]{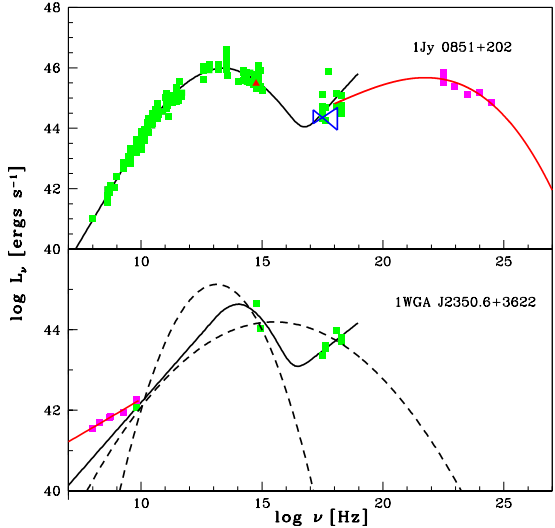}
  \caption{Examples of typical blazar SEDs in the master sample of
    flat-spectrum sources. Visual inspection of over 1300 SEDs was
    used to separate sources with unambiguous synchrotron peaks (top,
    653 sources total) from those with under-sampled SEDs which might
    be consistent with several possible peak locations (bottom). The
    dark line is the best-fit model (described in Appendix
    \ref{app:SEDfit}) for each case, while the dashed lines show other
    possible SED shapes consistent with the more sparsely-sampled SED.}
  \label{fig:goodandbad}
\end{figure}

%------------------Blazar Table-------------------------------------------------
%\LongTables
\begin{deluxetable*}{l c c c c c c c c c c}
\tabletypesize{\footnotesize}
%\rotate
\tablecolumns{11}
\tablewidth{0pt}
\tablecaption{TEX Combined Blazar Source List\label{table:allsources}}

\tablehead{
\colhead{Name} & \colhead{R.A.} & \colhead{Decl.} & \colhead{Opt.\tablenotemark{$\ddagger$}} &\colhead{z} & \colhead{log $\nu_{\mathrm{peak}}$} & \colhead{log $\mathrm{L}_{\mathrm{peak}}$} & \colhead{log $\mathrm{L}_{300}$} & \colhead{log $\mathrm{L}_{\mathrm{\mathrm{kin}}}$} & \colhead{Method} & \colhead{Ref.} \\
& (J2000) & (J2000) & & & (Hz) & (ergs s$^{-1}$) & (ergs s$^{-1}$) & (ergs s$^{-1}$) & & \\
(1) & (2) & (3) & (4) & (5) & (6) & (7) & (8) & (9) & (10) & (11) \\
}
\startdata
QSO B0003$-$06 & 00 06 13.89 & $-$06 23 35.33 & B & 0.347 & 13.16 & 46.22 & 41.68 & 44.21 & both & f,h \\
QSO B0016+73 & 00 19 45.73 & $+$73 27 30.07 & Q & 1.781 & 13.26 & 47.65 & 42.61 & 44.74 & both & f \\
QSO B0048$-$09 & 00 50 41.31 & $-$09 29 05.21 & B & 0.200 & 13.43 & 44.97 & 41.20 & 43.93 & both & b,d,f \\
4C 01.02 & 01 08 38.77 & $+$01 35 00.27 & Q & 2.099 & 13.00 & 47.71 & 44.57 & 45.85 & both & c,f,h \\
QSO B0109+224 & 01 12 05.82 & $+$22 44 38.78 & B & 0.265 & 13.87 & 45.57 & 40.33 & 43.44 & both & b,f,h \\
QSO B0118$-$272 & 01 20 31.66 & $-$27 01 24.65 & B & 0.557 & 14.04 & 46.13 & 42.38 & 44.61 & VLA & d \\
QSO B0133+47 & 01 36 58.58 & $+$47 51 29.30 & Q & 0.859 & 13.34 & 46.63 & 41.93 & 44.35 & VLA & f,h \\
QSO B0138$-$097 & 01 41 25.83 & $-$09 28 43.67 & B & 0.733 & 13.49 & 46.22 & 42.41 & 44.62 & VLA & d \\
4C 15.05 & 02 04 50.41 & $+$15 14 11.04 & Q & 0.405 & 12.39 & 45.15 & 41.39 & 44.04 & VLA & h \\
QSO B0212+73 & 02 17 30.82 & $+$73 49 32.55 & B & 2.367 & 13.38 & 47.20 & 42.30 & 44.56 & VLA & f,h \\
QSO B0215+015 & 02 17 48.95 & $+$01 44 49.69 & Q & 1.715 & 13.72 & 47.71 & 43.52 & 45.26 & VLA & b,f,h \\
QSO B0218+357 & 02 21 05.48 & $+$35 56 13.70 & Q & 0.960 & 12.86 & 45.91 & 43.09 & 45.01 & SD &  \\
3C 66A & 02 22 39.61 & $+$43 02 07.80 & B & 0.444 & 13.98 & 46.34 & 42.85 & 44.87 & both & b,e \\
\enddata
% }
\tablecomments{Table \ref{table:allsources} is published in its
  entirety in the electronic edition of the {\it Astrophysical
    Journal}. A portion is shown here for guidance regarding its form
  and content. Units of right ascension are hours, minutes, and
  seconds, and units of declination are degrees, arcminutes, and
  arcseconds.}  \tablenotetext{*}{Redshift estimated from optical
  magnitude} \tablenotetext{$\dagger$}{Spectral decomposition estimate
  of extended flux used instead of VLA estimate}
\tablenotetext{$\ddagger$}{Optical Spectral Type. B = BL Lac, Q = FSRQ.}
\tablenotetext{b}{\cite{ant85}} \tablenotetext{c}{\cite{mur93}}
\tablenotetext{i}{\cite{cas99_1jy_ext}}
\tablenotetext{e}{\cite{lan08}} \tablenotetext{f}{\cite{kha10}}
\tablenotetext{g}{\cite{con02_extflux}}
\tablenotetext{h}{\cite{coo07_mojave_iii}}
\tablenotetext{i}{\cite{cac04_class_testing_the_sequence}}
\end{deluxetable*}
%-----------------

%\input{table_blazars.txt}
To fit our model, we used a simulated annealing routine for the
parameter search \citep[][]{gof94}, and an algorithm based on a
combination of minimum square distance (in log $\mathrm{L}_{\nu}$)
between model and data, with the added constraint of reasonable
agreement with the X-ray spectral index where available, enforced by
increasing the score of the fit by a function of the angle between the
model and observed spectral slope.

We note that while simultaneous multi-wavelength observations give the
best `snapshot' view of an entire SED, such data is usually still
biased by being limited to a single epoch, as the variability of these
objects is well-known. The compilation of all available photometry,
generally collected over a long time-frame, sampling many states of
the SED, allows us to fit the model to the average SED of the source
which may be more appropriate in a population context, in which extreme
source states might confuse results.

\subsection{The TEX Sample of Blazars}
After fitting with the SED model, each source was subsequently
examined by eye, in order to identify ambiguous cases where the formal
best-fit model did not clearly represent the only reasonable SED
fit. In many cases these removed sources had few data points in the
critical region from IR to UV, lacked X-ray spectral indices, or had
significant host galaxy contamination in the optical.  A comparison of
an acceptable and a rejected source with their best SED fits is shown
in Figure \ref{fig:goodandbad}. A total of 653 sources remained after
this examination, in which we feel confident in the determination of
the synchrotron peak to within one order of magnitude.  (The rejected
sources are discussed along with the other under-sampled blazars in \S
\ref{sec:discards}). In total, 216 sources have robust estimates of
$\nu_{\mathrm{peak}}$, $\mathrm{L}_{\mathrm{peak}}$, and
$\mathrm{L}_{300}$; this set comprises the TEX (trusted extended)
sample. Of these, 129 have extended radio flux from spectral
decomposition, 87 were added using only VLA map data, and 90 have
both a spectral and VLA estimate.  For the remaining 437 blazars
comprising the UEX (unknown extended) sample, upper limits on
$\mathrm{L}_{300}$ (and subsequently $\mathrm{L}_{\mathrm{kin}}$) were
found by extrapolating from the luminosity at the lowest available
frequency using $\alpha$ = 1.2.

In Table \ref{table:allsources} we list the sources which have a
confirmed synchrotron peak and extended power measurement (TEX
sample). In columns 1 $-$ 5 we give the source name, position, optical
type, and redshift. Estimated redshifts are noted by an asterisk in
column 5. In columns 6 and 7 we give the best-fit synchrotron peak
frequency and luminosity, respectively. In columns 8 and 9 we give
$\mathrm{L}_{300}$ and the converted value of
$\mathrm{L}_{\mathrm{kin}}$, respectively. In column 10 we note the
method for the extended radio flux estimate: VLA, spectral
decomposition (SD), or both. We note those sources (19 total) in
which the spectral result is used instead of a VLA measurement by a
dagger in that column. In column 11 we give the references for VLA
data.  All values are in the rest-frame of the source.

% table 2 here
Because of the somewhat subjective requirement of an unambiguous
synchrotron peak, the TEX sample does not contain all the sources in
original list, and cannot be considered a complete sample. We focus
most of our discussion on qualitative analysis and positive findings
for our well-characterized sample. Relaxing the criteria for inclusion
in the sample risks inaccurate measurements of the three principle
quantities ($\nu_{\mathrm{peak}}$, $\mathrm{L}_{\mathrm{peak}}$, and
$\mathrm{L}_{\mathrm{kin}}$) and could lead to incorrect conclusions. See
\S \ref{section:biases} for a discussion of the selection effects of
our master sample, the nature of the rejected sources, and the
possible impact on our conclusions.

\subsection{Radio Galaxies}
As the jets isolated in radio galaxies are understood to be the
less-beamed counterparts to blazars, one way to understand the
relationship of the synchrotron peak and beaming in the envelope is to
see where the synchrotron peak falls for these jets. In the context of
the presumed blazar sequence, these `hidden blazars' should indicate
whether the sequence is preserved in misaligned jets. We take
advantage of recent work to isolate the radio, IR, optical, and X-ray
core emission from the jet in radio galaxies using high-resolution
mapping to place as many radio galaxies as possible on the
$\nu_{\mathrm{peak}}$ - $\mathrm{L}_{\mathrm{peak}}$ plane
\citep[][]{chi99,cap00,chi02,cap02,har03,tru03,bal06,mas10,but11}. We
examine the 45 radio galaxies with extended radio flux measurements,
and at least one radio, optical, and X-ray measurement of the nuclear
emission available in the literature. The $\mathrm{L}_{300}$ for all
radio galaxies was found from a parabolic fit\footnote{For fitting the
  steep component in radio galaxies, a log-parabolic model is more
  appropriate than a power-law due to the much greater range of the
  spectrum visible, which usually has some curvature in $\nu
  F_{\nu}$.} to the steep, low-frequency radio emission and
subsequently converted to $\mathrm{L}_{\mathrm{kin}}$ as in Equation
\ref{eq:qjet}. In all cases where available, VLA map-derived estimates
at 1.4 GHz were in agreement with the spectral fits.

Fourteen radio galaxies are nearby, well-known sources, with core
measurements at multiple frequencies which allow fitting of the
synchrotron SED with the same parameterization as for the
blazars. This set includes seven of the Fermi-detected
\emph{misaligned} AGN \citep{abd09_fermi_misaligned_agn}.

%------------------RG Table-------------------------------------------------
%\LongTables
\tabletypesize{\footnotesize}
\begin{deluxetable*}{l c c c c c c c c}
%\rotate
\tablecolumns{9}
\tablewidth{0pt}
\tablecaption{Radio Galaxies\label{table:radiogalaxies}}

\tablehead{
\colhead{Name} & \colhead{R.A.} & \colhead{Decl.} & \colhead{Morph.\tablenotemark{$\ddagger$}} & \colhead{z} & \colhead{log $\nu_{\mathrm{peak}}$} & \colhead{log $\mathrm{L}_{\mathrm{peak}}$} & \colhead{log $\mathrm{L}_{300}$} & \colhead{$\mathrm{L}_{\mathrm{kin}}$}  \\
& (J2000) & (J2000) & & & (Hz) & (ergs s$^{-1}$) & (ergs s$^{-1}$) & (ergs s$^{-1}$) \\
(1) & (2) & (3) & (4) & (5) & (6) & (7) & (8) & (9) \\
}
\startdata
3C 380\tablenotemark{$\dagger$} & 18 29 31.77 & $+$48 44 46.17 & 2 & 0.69 & 12.95 & 45.68 & 44.06 & 45.56 \\
          3C 17 & 00 38 20.52 & -02 07 40.8 & 2 & 0.220 & 13.05$^{+0.30}_{-0.55}$ & 45.35 & 42.82 & 44.86 \\
          3C 18 & 00 40 50.35 & +10 03 23.14 & 2 & 0.188 & 12.83$^{+0.32}_{-0.52}$ & 44.79 & 42.55 & 44.70 \\
          3C 29 & 00 57 34.92 & -01 23 27.89 & 1 & 0.045 & 12.94$^{+0.39}_{-0.37}$ & 44.01 & 41.34 & 44.02 \\
        NGC 315 & 00 57 48.88 & +30 21 08.81 & 1 & 0.016 & 13.96$^{+0.34}_{-0.65}$ & 43.94 & 39.78 & 43.13 \\
          3C 31 & 01 07 24.93 & +32 24 45.20 & 1 & 0.017 & 13.09$^{+0.50}_{-0.41}$ & 43.53 & 40.79 & 43.70 \\
        3C 33.1 & 01 09 44.26 & +73 11 57.17 & 2 & 0.181 & 12.99$^{+0.36}_{-0.25}$ & 44.30 & 42.42 & 44.63 \\
 3C 66B\tablenotemark{$\dagger$} & 02 23 11.41 & $+$42 59 31.38 & 1 & 0.02 & 13.36 & 42.09 & 40.73 & 43.67 \\
 3C 78\tablenotemark{$\dagger$} & 03 08 26.22 & $+$04 06 39.30 & 1 & 0.03 & 12.99 & 44.02 & 40.94 & 43.79 \\
        3C 83.1B & 03 18 15.73 & +41 51 27.38 & 1 & 0.025 & 13.09$^{+0.32}_{-0.62}$ & 43.40 & 40.93 & 43.78 \\
 3C 84\tablenotemark{$\dagger$} & 03 19 48.16 & $+$41 30 42.10 & 1 & 0.02 & 13.46 & 43.09 & 40.16 & 43.34 \\
3C 111\tablenotemark{$\dagger$} & 04 18 21.27 & $+$38 01 35.80 & 2 & 0.05 & 13.28 & 43.29 & 41.41 & 44.05 \\
         3C 133 & 05 02 58.50 & +25 16 24.0 & 2 & 0.278 & 13.04$^{+0.23}_{-0.22}$ & 45.08 & 42.98 & 44.95 \\
         3C 135 & 05 14 08.34 & +00 56 31.63 & 2 & 0.127 & 13.05$^{+0.35}_{-0.36}$ & 43.79 & 42.04 & 44.41 \\
         3C 165 & 06 43 06.70 & +23 19 00.3 & 2 & 0.296 & 13.04$^{+0.23}_{-1.04}$ & 44.33 & 42.80 & 44.84 \\
         3C 171 & 06 55 14.87 & +54 08 59.4 & 2 & 0.238 & 13.04$^{+0.22}_{-0.22}$ & 43.89 & 42.73 & 44.81 \\
3C 189\tablenotemark{$\dagger$} & 07 58 28.10 & $+$37 47 11.80 & 1 & 0.04 & 13.44 & 42.34 & 40.88 & 43.75 \\
         3C 264 & 11 45 05.00 & +19 36 22.74 & 1 & 0.022 & 12.60$^{+0.52}_{-0.32}$ & 43.77 & 40.72 & 43.66 \\
         3C 270 & 12 19 23.21 & +05 49 29.69 & 1 & 0.007 & 13.16$^{+0.31}_{-1.16}$ & 43.33 & 40.09 & 43.30 \\
       NGC 4278 & 12 20 06.82 & +29 16 50.71 & 1 & 0.002 & 12.96$^{+0.39}_{-0.67}$ & 42.55 & 37.29 & 41.70 \\
  M 84\tablenotemark{$\dagger$} & 12 25 03.74 & $+$12 53 13.13 & 1 & 0.00 & 12.98 & 41.19 & 39.04 & 42.70 \\
  M 87\tablenotemark{$\dagger$} & 12 30 49.42 & $+$12 23 28.04 & 1 & 0.00 & 13.01 & 41.58 & 40.95 & 43.79 \\
Cen A\tablenotemark{$\dagger$} & 13 25 27.61 & $-$43 01 08.80 & 1 & 0.00 & 12.72 & 41.47 & 40.21 & 43.37 \\
       3C 287.1 & 13 32 53.27 & +02 00 45.79 & 2 & 0.216 & 13.05$^{+0.26}_{-1.05}$ & 45.21 & 42.38 & 44.61 \\
3C 296\tablenotemark{$\dagger$} & 14 16 52.92 & $+$10 48 26.43 & 1 & 0.02 & 12.99 & 41.24 & 40.57 & 43.57 \\
         3C 300 & 14 23 01.00 & +19 35 17.0 & 2 & 0.272 & 12.73$^{+0.40}_{-0.45}$ & 44.32 & 42.85 & 44.87 \\
3C 317\tablenotemark{$\dagger$} & 15 16 44.48 & $+$07 01 18.00 & 1 & 0.03 & 12.83 & 42.95 & 41.34 & 44.01 \\
         3C 321 & 15 31 43.47 & +24 04 18.93 & 2 & 0.096 & 13.27$^{+0.60}_{-0.48}$ & 44.19 & 41.86 & 44.31 \\
B2 1553+24 & 15 56 03.87 & +24 26 52.70 & 1 & 0.043 & 13.11$^{+0.31}_{-1.11}$ & 43.79 & 40.48 & 43.52 \\
3C 338\tablenotemark{$\dagger$} & 16 28 38.27 & $+$39 33 04.97 & 1 & 0.03 & 13.12 & 41.44 & 41.18 & 43.92 \\
NGC 6251\tablenotemark{$\dagger$} & 16 32 31.96 & $+$82 32 16.39 & 1 & 0.02 & 12.79 & 43.25 & 40.68 & 43.64 \\
3C 346\tablenotemark{$\dagger$} & 16 43 48.62 & $+$17 15 48.97 & 1 & 0.16 & 13.46 & 43.49 & 42.25 & 44.53 \\
         Her A & 16 51 08.14 & +04 59 33.32 & 2 & 0.155 & 13.04$^{+0.22}_{-0.22}$ & 44.08 & 43.60 & 45.30 \\
         3C 349 & 16 59 32.00 & +47 02 16.0 & 2 & 0.205 & 12.94$^{+0.25}_{-0.45}$ & 44.39 & 42.49 & 44.67 \\
         3C 388 & 18 44 02.41 & +45 33 29.81 & 2 & 0.091 & 13.03$^{+0.46}_{-0.39}$ & 44.32 & 42.00 & 44.39 \\
         3C 442A & 22 14 45.00 & +13 50 47.66 & 1 & 0.026 & 12.98$^{+0.53}_{-0.65}$ & 42.74 & 40.77 & 43.69 \\
         3C 449 & 22 31 20.62 & +39 21 29.81 & 1 & 0.017 & 13.18$^{+0.51}_{-0.57}$ & 43.27 & 40.17 & 43.35 \\
        B2 2236+35 & 22 38 29.46 & +35 19 40.01 & 1 & 0.027 & 13.04$^{+0.23}_{-0.22}$ & 43.03 & 39.62 & 43.03 \\
         3C 452 & 22 45 48.78 & +39 41 15.36 & 2 & 0.081 & 13.02$^{+0.23}_{-1.02}$ & 44.48 & 43.34 & 45.15 \\
         3C 460 & 23 21 28.50 & +23 46 48.0 & 2 & 0.268 & 13.04$^{+0.24}_{-1.04}$ & 44.52 & 42.52 & 44.69 \\
         3C 465 & 23 38 29.39 & +27 01 53.52 & 1 & 0.031 & 13.07$^{+0.38}_{-0.61}$ & 44.07 & 41.12 & 43.89 \\
\enddata
% }
\tablecomments{Units of right ascension are hours, minutes, and seconds, and units of declination are degrees, arcminutes, and arcseconds.}
\tablenotetext{$\dagger$}{Synchrotron peak estimated from blazar model fit to core fluxes}
\tablenotetext{$\ddagger$}{Morphological Type. 1 = FR I, 2 = FR II.}
\end{deluxetable*}
%-----------------

%\renewcommand{\tabcolsep}{1cm}
%\input{table_rg.txt}

For the remaining 31 with sparser SED sampling, the synchrotron
$\nu_{\mathrm{peak}}$ was estimated using a non-parametric likelihood
estimator.  The joint distribution of $\mathrm{R}_{\mathrm{CE}}$,
$\nu_{\mathrm{peak}}$, and the SED colors\footnote{Spectral indices
  are defined at 1.4 GHz, 5000 \AA, and 1 keV.} $\alpha_{ro}$,
$\alpha_{ox}$ was calculated for all the sources in our TEX, UEX, and
fitted radio galaxy samples. From this, a conditional, one-dimensional
density on $\nu_{\mathrm{peak}}$ can be found by supplying the other
three observables. The value of the peak was taken as the maximum of
this distribution, and we also report the 20\% and 80\% quartiles as
the sampling error.

It is straightforward to estimate the peak luminosity from the radio
luminosity at 5 GHz, with which it has a linear correlation
(correlation coefficient r = 0.85).  The linear fit (with 1-$\sigma$
error) is
\begin{equation}
  \begin{split}
\mbox{log }\mathrm{L}_{\mathrm{peak}} = \mbox{0.61}\left(\pm\mbox{0.01}\right)&\,\left(\mbox{log }\mathrm{L}_{\mathrm{5 GHz}}-43\right)\\
&+ \mbox{45.68}\pm\left(\mbox{0.02}\right)\mbox{  [ergs s}^{-1}\mbox{]}\\
\end{split}
\end{equation}  The 90\% confidence
interval, assuming a normal error distribution, corresponds to an
uncertainty of $\Delta$(log $\mathrm{L}_{\mathrm{peak}}$) $\simeq$ 0.8.

The most likely peak frequency found by our statistical model for all
radio galaxies is at or below $10^{14}$ Hz, with the exception of four
FR II sources with moderate jet powers ($\mathrm{L}_{\mathrm{kin}}$
$\sim$ $10^{44.5}$ ergs s$^{-1}$) which display extreme X-ray
luminosities.  These sources include the broad-lined radio galaxies
(BLRG) 3C 390.3 (which has been previously reported to have an
LBL-like SED by \citealt{gho95_3c3903}) and 3C 332, as well as two
lesser-studied narrow line radio galaxies, 3C 197.1 and 3C 184.1.  The
two BLRG have been detected with Swift and BeppoSAX in the hard X-ray
\citep{gra06,cus10}. All four sources were excluded from the final
sample, as the X-ray is likely either thermal or from a strong IC
component, and without knowledge of the synchrotron X-ray emission,
any peak estimate would be practically arbitrary.

The final list of 41 radio galaxies is given in Table
\ref{table:radiogalaxies}. Column descriptions are as in Table
\ref{table:allsources}.

\section{RESULTS}
\label{results}

\subsection{The Blazar Envelope}
\begin{figure}
  \includegraphics[width=0.99\linewidth]{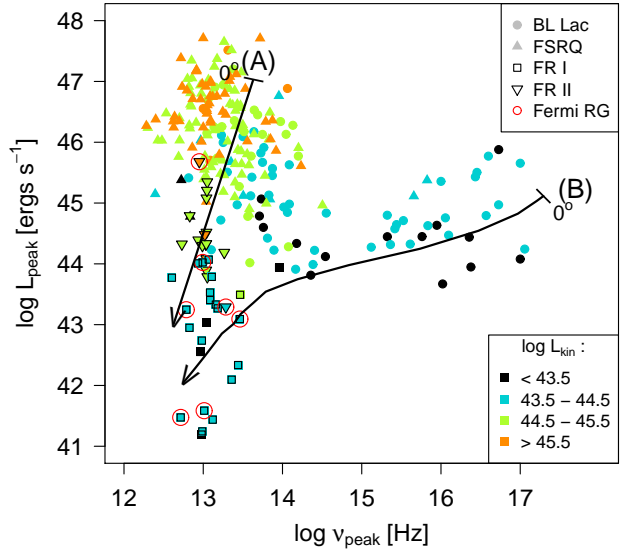}
  \caption{The blazar sequence, which originally showed an
    anti-correlation between synchrotron $\mathrm{L}_{\mathrm{peak}}$
    and $\nu_{\mathrm{peak}}$ has been expanded into an `envelope'
    with the addition of new observations and radio galaxies. BL Lacs
    are shown as filled circles, FSRQ as filled triangles, and radio
    galaxies as squares (FR I) and inverted triangles (FR II).  Color
    indicates the jet kinetic power ($\mathrm{L}_{\mathrm{kin}}$ in
    ergs s$^{-1}$), as estimated from extended radio flux
    measurements. Track (A) shows the path of a synchrotron peak for a
    single-component jet, and (B) for a decelerating jet of the type
    hypothesized to exist in FR I sources. The fully-aligned limit for each (0
    degrees) is shown as marked, with the arrow direction indicating
    the movement of the model source as it is misaligned.}
  \label{fig:envelope}
\end{figure}

The TEX sample of 216 blazars, and the 41 radio galaxies are shown in
the plane of $\mathrm{L}_{\mathrm{peak}}$ versus $\nu_{\mathrm{peak}}$
in Figure \ref{fig:envelope}. Blazars are represented as circles (BL
Lacs) and triangles (FSRQ). The power of the source
($\mathrm{L}_{\mathrm{kin}}$) is shown by color in bins of one
decade. Examining the location of blazars, there is an obvious lack of
sources with both high jet powers \emph{and} synchrotron peak
frequencies above $10^{15}$ Hz, though a few of the ISP/HSP sources
have quasar-like spectra.  The population at low peak frequencies,
alternately, is highly mixed in terms of kinetic power (with many
low-power sources coincident with high-power sources), spectral type
(with both BL Lac and quasar spectra present) and the spread of peak
luminosities, which ranges from $10^{44}$ to $10^{47.5}$ ergs
s$^{-1}$.

The 41 radio galaxies are shown in Figure \ref{fig:envelope} as
squares (FR I) and inverted triangles (FR II). Those detected by {\sl
  Fermi} \citep{abd09_fermi_misaligned_agn} are circled in red.  For
our radio galaxy sample, we note that the jet power does not
apparently exert much influence on their synchrotron peak frequency,
and it appears that all radio galaxies in our sample exhibit jets with
peaks below the optical-UV range in frequency. Further, no evidence is
found for any conservation of a `blazar sequence' within the radio
galaxy jets shown here. If, as has been suggested, different emission
regions in the less powerful jets dominate at different orientation
angles \citep{chi00, geo03,ghi05} it may simply be that the `sequence'
behavior (in which jet power regulates the synchrotron peak) pertains
to the fast parts of the jet which are no longer visible in highly
misaligned radio galaxies.

It becomes interesting to ask the question of what type of orientation
tracks are expected on the
$\mathrm{L}_{\mathrm{peak}}$-$\nu_{\mathrm{peak}}$ plane if we assume
that powerful sources do not decelerate, while weak sources require
velocity gradients. The former assumption is supported by the fact
that the Lorentz factors required to model the blazar SED of these
sources ($\Gamma\sim 10-20$) are similar with those deduced from
following VLBI superluminal components \cite[e.g.][]{jor01}, while the
latter (as we discussed in \S \ref{velocitygrad}) is a result of our
need to ({\sl i}) explain the discrepancy between the high Lorentz
factors required by the fast variability, pair production opacity, and
SED modeling for the sub-pc scale jet and the lack of significantly
superluminal motions at the VLBI scales and ({\sl ii}) satisfy the
unification between low power blazars and FR I radio galaxies.

Two tracks are shown in Figure \ref{fig:envelope}.  These illustrative
tracks do not intend to simulate any particular source, and are shown
to simply demonstrate how sources that start with the jet pointing
toward us will appear as the jet orientation angle increases. The
first, labeled (A), follows the misalignment of a powerful source
assuming a single jet Lorentz factor $\Gamma= 15$.  This track follows
the expected 1:4 ratio of decreasing (log) frequency to (log)
luminosity of the peak as the source is decreasingly beamed
\citep[see, \emph{e.g.},][]{lin85}.  The second track, labeled (B), is
for a weak source and shows the change in peak location for one
possible model of a decelerating jet, as described in \cite{geo03}. In
this particular example, the Lorentz factor of the flow decreases
continuously from $\Gamma_{max}=15$ to $\Gamma_{min}=3$ in a distance
of $5 \times 10^{17}$ cm, and the emitted spectrum is calculated along
the jet and integrated spatially.  The track exhibits the qualitative
characteristics (\emph{e.g.}, more horizontal movement) not only
consequent of the decelerating model, but of \emph{any model} with
velocity gradients in which the most energetic electrons are injected
in the fast part of the flow, such as the spine-sheath model of
\cite{ghi05}.  Such models, as can be seen in Figure
\ref{fig:envelope}, exhibit a shallow decrease of
$\mathrm{L}_{\mathrm{peak}}$ and a sharp decrease of
$\mathrm{\nu}_{\mathrm{peak}}$ because the opening angle of the
beaming cone is smaller for higher frequency emission; as the angle to
the line of sight increases, progressively lower frequency emission is
beamed away from the observer. Only for angles $\gtrsim
1/\Gamma_{min}$, do we expect the beaming track to start resembling
that of a constant Lorentz factor flow and
$\mathrm{L}_{\mathrm{peak}}$ to start dropping faster with decreasing
$\mathrm{\nu}_{\mathrm{peak}}$.  This model predicts that the
low-power HSP blazars will follow a more horizontal path from the most
beamed blazars to the misaligned radio galaxies. Interestingly, though
each model originates in very different parts of the plane at low
viewing angles, they coincide at higher orientation angles
(\emph{i.e.}, when seen as radio galaxy core emission).

\begin{figure}
 \includegraphics[width=0.99\linewidth]{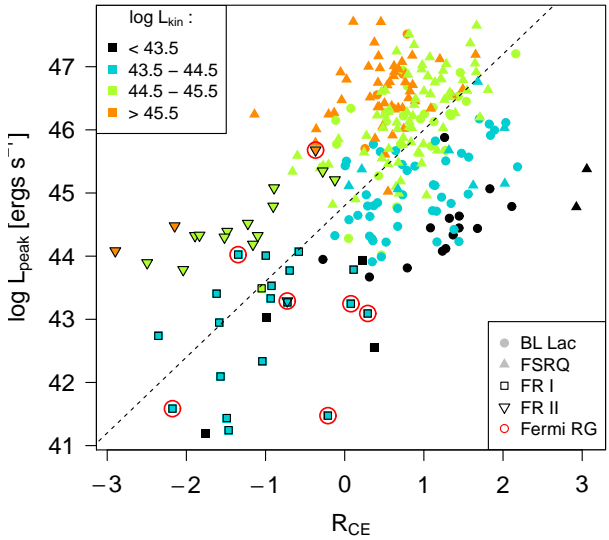}
  \caption{Core dominance ($\mathrm{R}_{\mathrm{CE}}$) versus
    synchrotron peak luminosity for BL Lacs, FSRQ, and radio galaxies
    in our sample.  Color indicates the kinetic jet power
    ($\mathrm{L}_{\mathrm{kin}}$ in ergs s$^{-1}$), as calculated from
    extended radio flux measurements, as in Figure
    \ref{fig:envelope}. The dotted line has a slope equal to 1.2, the
    OLS bisector slope found for the whole sample. This slope is
    consistent with the expected relation between
    $\mathrm{L}_{\mathrm{peak}}$ and $\mathrm{L}_\mathrm{R}$ due to
    beaming effects.}
  \label{fig:cdlp}
\end{figure}

\subsection{Radio Core Dominance}
\label{sec:CD}
In order to investigate the effect of viewing angle in populating the
blazar envelope, it is necessary to have a way to measure or
approximate the degree of misalignment in individual sources.  The
most readily available measure is the radio core dominance
$\mathrm{R}_{\mathrm{CE}}$; typical values range from $-$3 (for the
most off-axis radio galaxies) to $\sim$ 3 (highly beamed quasars). As
the extended emission is non-variable and unaffected by beaming, the
dependence of $\mathrm{R}_{\mathrm{CE}}$ on the orientation angle
$\theta$ is directly due to the increased Doppler boosting factor at
low angles. The Doppler factor $\delta$ is given by
\begin{equation}
\delta = \frac{1}{\Gamma}\frac{1}{\left(1 - \beta cos\theta\right)}
\end{equation}
where $\Gamma$ is the Lorentz factor of the jet and $\beta$ = v/c.  The
Doppler-boosted luminosity for any part of the jet SED is a function
of $\delta$,
\begin{equation}
\mathrm{L} = \mathrm{L}_0\delta^{p}
\label{eq:dop}
\end{equation}
where $\mathrm{L}^\prime$ is the unbeamed jet luminosity and p = 2 +
$\alpha$ to 3 + $\alpha$ depending on the geometry of the jet model,
with $\alpha$ equal to the energy spectral index at the frequency of
interest \citep[see][]{urr95}. $\mathrm{R}_{\mathrm{CE}}$ is thus a
monotonically decreasing function of $\theta$, though the exact shape
is dependent on $\Gamma$, becoming steeper as $\theta$ increases. This
does create a difficulty in comparing $\mathrm{R}_{\mathrm{CE}}$
values directly for sources which may have very different bulk
$\Gamma$ values (such as, perhaps, very high and low power
sources). However, if we assume that sources of the same jet power are
similar in $\Gamma$, then we can compare at least these directly, and
in general a source with a lower $\mathrm{R}_{\mathrm{CE}}$ value will
be more misaligned.

\begin{figure}
    \includegraphics[width=0.99\linewidth]{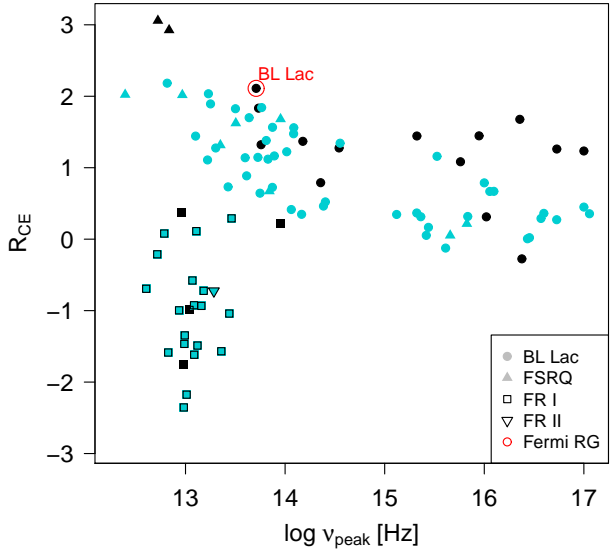}

    \caption{Core dominance ($\mathrm{R}_{\mathrm{CE}}$) versus the
      synchrotron peak frequency for all sources with
      $\mathrm{L}_{\mathrm{kin}}$ below $10^{44.5}$
      ($\mathrm{L}_{300}$ below $\sim$ $10^{42}$) ergs s$^{-1}$. While all
      low (kinetic) power radio galaxies have lower core dominance
      than blazars of similar power, there are apparently two
      locations of more aligned low-power blazars $-$ one at low peak
      frequencies, with relatively higher $\mathrm{R}_{\mathrm{CE}}$
      values (up to 3), and another at high peak frequencies, with
      slightly lower $\mathrm{R}_{\mathrm{CE}}$ values ($\sim$
      1$-$1.5). While nearly all sources at high peaks ($> 10^{15}$ Hz) are
      BL Lacs (\emph{i.e.}, HBLs), the low-peaking aligned sources
      include a mixture of FSRQ and BL Lac spectral types. Color scale
      is the same as used in Figure \ref{fig:cdlp}.}
  \label{fig:cdvp}
\end{figure}

In order to investigate further how the sources of similar jet power
are related in Figure \ref{fig:envelope}, it would be helpful to add
the core dominance.  We thus examine how the core dominance changes
with peak luminosity and frequency separately in Figures
\ref{fig:cdlp} and \ref{fig:cdvp}, still continuing to bin by the jet
kinetic power as before.  This choice is partly because it is
practically difficult to add a 4th parameter in Figure
\ref{fig:envelope}, and also because, as suggested in Figure
\ref{fig:cdlp}, the core dominance may have a different maximum (i.e.,
normalization), depending on the jet power, such that high-power
sources have a lower maximum $\mathrm{R}_\mathrm{CE}$ than that of the
lowest power bin, making it difficult to compare sources at different
powers directly. This can be understood if we make the assumption that
unbeamed radio power scales from the overall jet power as
$\mathrm{L}_\mathrm{R}^\prime = \epsilon\mathrm{L}_\mathrm{kin}$
implying that jet radiative efficiency does not change much as the jet
power increases. As we know
$\mathrm{L}_\mathrm{kin}$=$\kappa\mathrm{L}_\mathrm{ext}^\beta$
(taking $\beta\sim$ 0.6, see \S \ref{sec:jetpower}), this leads to the
relation
\begin{equation}
\mbox{max }\mathrm{R}_\mathrm{CE}=\left(1-1/\beta\right)\mbox{log }
\mathrm{L}_\mathrm{kin} + p\times\mbox{log}\left(2\Gamma\right) + c_1
\end{equation}
\noindent
where we have let $\delta_\mathrm{max}$=2$\Gamma$, $c_1$ =
log($\epsilon\kappa^{1/\beta}$), and (1-1/$\beta$) $\sim$
$-$2/3. Assuming the last term is roughly constant, the maximum value of
$\mathrm{R}_\mathrm{CE}$ is thus \emph{expected to decrease with
  increasing $\mathrm{L}_\mathrm{kin}$, even assuming reasonably different
  $\Gamma$ values for high and low-power sources}.  For example, taking p=2,
and $\Gamma$=15 for a jet of $\mathrm{L}_\mathrm{kin}$ = 10$^{46}$
predicts a max $\mathrm{R}_\mathrm{CE}$ about 1.25 lower than for a jet
with $\mathrm{L}_\mathrm{kin}$ = 10$^{42}$ and $\Gamma$=3.

It is also expected that sources will separate by jet kinetic power in
Figure \ref{fig:cdlp}, as following from the above equations, we have
for the expected relation between $\mathrm{R}_\mathrm{CE}$ and
$\mathrm{L}_\mathrm{peak}$:

\begin{equation}
  \mathrm{R}_\mathrm{CE} = \left(\frac{2+\alpha_R}{3}\right)\mbox{log
  }\mathrm{L}_\mathrm{peak} + \left(1-\frac{1}{\beta}\right)\mbox{log
  }\mathrm{L}_\mathrm{kin} + c_2
\label{eq:lpcd}
\end{equation}

It is clear that binning by $\mathrm{L}_\mathrm{kin}$ will indeed
induce the `striping' in that figure through the second term, and that
the slope in Figure \ref{fig:cdlp} should be about 1.2 - 1.5 depending
on the value of $\alpha_R$ (where we have assumed p = 2 + $\alpha$;
see derivation in appendix \ref{app:deriv}). However, scatter could be
induced by the remaining term, $c_2$ = log($\epsilon
\kappa^{1/\beta}$) $-$ (2+$\alpha_R$)log$\mathrm{L}_\mathrm{peak}^\prime$/3
which is clearly determined by the relation between the unbeamed
luminosity of a source at a given $\mathrm{L}_\mathrm{kin}$.  If this
quantity is not roughly constant with respect to
$\mathrm{L}_\mathrm{kin}$, Figure \ref{fig:cdlp} would be reduced to a
multi-color scatter plot and the striping would be lost or degraded.
It is also interesting to note that, like in Figure \ref{fig:xcross},
there is a fairly uniform, sharp transition from blazars to radio
galaxies in Figure \ref{fig:cdlp}. These arguments again suggest that
beaming has a dominant effect on the observed (core) luminosity,
rather than being due to a range of unbeamed core luminosities
($\mathrm{L}_\mathrm{R}^\prime$, $\mathrm{L}_\mathrm{peak}^\prime$) for a given
$\mathrm{L}_\mathrm{kin}$.

A mild decrease in synchrotron peak frequency with increasing angle to
the line of sight is expected in the simplest case of a jet dominated
by a single Lorentz factor. The change in peak frequency as core
dominance decreases is not significant for our sample of sources with
$\mathrm{L}_{\mathrm{kin}}$ greater than 10$^{44.5}$ ergs s$^{-1}$,
all of which have low synchrotron peak frequencies (these sources are
therefore not shown in Figure \ref{fig:cdvp}). For low-power sources
(less than 10$^{44.5}$ ergs s$^{-1}$), however, there is a surprising
pattern in the $\mathrm{R}_{\mathrm{CE}}-\nu_{\mathrm{peak}}$
plane. As shown in Figure \ref{fig:cdvp}, there appears to be a
clustering of sources with low peak frequencies and yet high core
dominance, where $\mathrm{R}_{\mathrm{CE}}$ reaches up to $\sim$3, and
another location at high peak frequencies where
$\mathrm{R}_{\mathrm{CE}}$ reaches $\sim$1$-$1.5. Either locus could
conceivably be connected via a misalignment track to the radio
galaxies at lower left, but the source distribution suggests that
there is not a smooth continuum between LSP and HSP sources. We expect
a similar monotonic relationship between $\mathrm{R}_{\mathrm{CE}}$
and $\nu_{\mathrm{peak}}$ as was derived above for the peak
luminosity, but this is clearly not consistent with what is shown in
Figure \ref{fig:cdvp}, at least if sources at low power consist of a
single population.

In some cases the high core dominance values for the LSP sources at
low powers could be due to systematic underestimation of
$\mathrm{L}_{300}$ for these objects because of known problems with
VLA measurements (incorrect modeling or use of an array configuration
which might miss extended flux).  However this would require nearly
every single source to be over-estimated by an order of magnitude or
more, and does not explain those LSP sources for which the steep
spectrum is actually seen and matches the VLA estimate (such as BL Lac
itself, noted in Figure \ref{fig:cdvp}).

\section{DISCUSSION}
\label{discussion}
\subsection{The Blazar Envelope}

\cite{ghi98} were able to reproduce the observed anti-correlations of
the \cite{fos98} blazar sequence using a simple one-zone leptonic jet
model. They suggested that the increased radiative cooling of the
electrons of high-power blazars lead to a particle energy distribution
with a break at lower energies (leading to the lower peak
frequency). In support of the scenario of a continuous spectral
sequence, our results in Figure \ref{fig:envelope} clearly show a
decreasing upper limit on $\mathrm{L}_{\mathrm{peak}}$ as
$\nu_{\mathrm{peak}}$ increases. That sources near this limit are
highly aligned, and those falling below are progressively misaligned
is supported by the strong correlation between core dominance and peak
luminosity shown in Figure \ref{fig:cdlp} (which we note is much
harder to appreciate without separating sources according to their
radio galaxy or blazar classification and by their jet kinetic power).
However, this simple scenario does not provide an explanation for the
rather noticeable `L' shape to the source distribution in Figure
\ref{fig:envelope} or the wide range of synchrotron peak frequencies
for sources of similar low jet powers.  While the former might be the
result of missing sources at intermediate peak frequencies, the latter
is a positive finding that cannot be explained away as a selection
effect.

\subsubsection{Jet Structures in the Envelope}
\label{sec:jetstruc}

\cite{ghi08_new_perspective} revisited the theoretical blazar
sequence, linking the power of the jet and the SED to two fundamental
parameters of the accretion disk $-$ the black hole mass and the
accretion rate. Following \cite{narayan97} they assumed that sources
with an accretion power ($L_{acc}$) lower than a critical value
$L_{cr}=L_{Edd} \dot{m}_{cr}$ are radiatively inefficient accretors
with weak or absent broad line region, while sources with
$L_{acc}>L_{cr}$ are characterized by efficiently radiating accretion
disks with strong big blue bumps and powerful broad line emission
(where $L_{Edd}$ is the Eddington luminosity and $\dot{m}_{cr}\sim3
\times 10^{-3}-10^{-2} $ a critical mass accretion rate in units of
the Eddington accretion rate).

As discussed in \S \ref{results} we see in our results the
manifestation of two populations distinguished by typical jet
structure.  The first population have non-decelerating jets
characterized by a single Lorentz factor.  These sources do not
contain a high-peaking synchrotron component and the increase in peak
frequency as the jet is aligned is modest, leading to a concentration
of blazars with synchrotron peak frequencies $\sim 10^{14}$ Hz even
when fully aligned. The sources in this population include all of the
more powerful blazars ($\mathrm{L}_{\mathrm{kin}}$ $>10^{44.5}$ ergs
s$^{-1}$) and many sources at lower powers. We include the latter
because of the high core dominance of many low-power, low-peaking
blazars in Figure \ref{fig:cdvp}. The second population is only
comprised of sources with $\mathrm{L}_\mathrm{kin}$ below $\sim
10^{44.5}$ ergs s$^{-1}$ and have complex, decelerating jets. These
lower-power sources exhibit a high frequency-peaking synchrotron
component when aligned.

The non-decelerating, low-peaking (`strong' jet) sources and
decelerating, high-peaking (`weak' jet) sources form distinct
populations which follow qualitatively different de-beaming paths in
$\mathrm{L}_{\mathrm{peak}}-\nu_{\mathrm{peak}}$ space, exampled by
track (A) and (B) in Figure \ref{fig:envelope}, respectively.  The two
population scenario explains the appearance of an L-shaped upper
boundary as a natural consequence of the vertical and horizontal paths
by sources through the
$\mathrm{L}_{\mathrm{peak}}-\nu_{\mathrm{peak}}$ plane. It also
explains the appearance of Figure \ref{fig:cdvp}, as low power sources
may be either strong or weak, producing considerable spread in
possible $\nu_{\mathrm{peak}}$ values as $\mathrm{R}_{\mathrm{CE}}$
increases. 

It is tempting to associate those weak sources (having
$\mathrm{L}_{\mathrm{kin}}$ $<10^{44.5}$ erg s$^{-1}$) that we argue
are characterized by velocity profiles (track B in Figure
\ref{fig:envelope}) with sources having $L_{acc}<L_{cr}$. If we assume
that $L_{kin}\lesssim L_{acc} $ \citep[e.g.][]{raw91_nature}, this
implies that the accretion power of these systems reaches at least up
to $L_{acc}\sim 10^{44.5}$ erg s$^{-1}$, which for $\dot{m}_{cr}\sim3
\times 10^{-3} $ corresponds to a black hole mass of $\sim 10^9 $
M$_\odot$, in agreement with black hole mass estimates \cite[e.g.,
][]{bar02}. Similarly, sources with $\mathrm{L}_{\mathrm{kin}}$
$>10^{44.5}$ erg s$^{-1}$, which are all of the powerful strong jet
type (track A in Figure \ref{fig:envelope}), must have
$L_{acc}>L_{cr}$ or equivalently $\dot{m}>\dot{m}_{cr}\sim3 \times
10^{-3}$, assuming that black hole masses do not significantly exceed
$\sim 10^9 $ M$_\odot$.
 
In this framework, the low jet power sources
($\mathrm{L}_{\mathrm{kin}}$ $<10^{44.5}$ erg s$^{-1}$) that yet seem
to have a strong jet may be explained as sources with relatively
smaller black hole masses, such that although their jets are low
power, the mass accretion rate required to sustain them is
$\dot{m}>\dot{m}_{cr}$. Consider a jet with
$\mathrm{L}_{\mathrm{kin}}=10^{44}$ erg s$^{-1}$ and black hole mass $
10^8 $ M$_\odot$. From the above, it is clear that the accretion rate
(in units of 10$^{-3}$) $\dot{m}_{-3}$ =
0.7 $\mathrm{L}_\mathrm{kin,44}$ $\mathrm{M}_9^{-1}$ $\sim$ 7, exceeds the
critical value placing this relatively low power jet in the family
of `strong' jets.

\subsubsection{Optical Spectral Type and the Envelope}

While the original observational sequence of \cite{fos98} implied a
transition from quasars to line-less BL Lacs along a sequence from
high to low bolometric luminosity, a firm distinction is not a
necessary part of the theoretical scheme for a continuous blazar
sequence. \cite{ghi08_new_perspective} explain the possibility of blue
FSRQ, formed when the dissipation radius for the jet exceeds the
radius of the broad line region, as well as lower-power FSRQ formed
when an efficient accretion forms around a relatively small black
hole, (the explanation we proposed in \S \ref{sec:jetstruc} is along
the same lines). However, the traditional divide between BL Lacs and
FSRQ, at an equivalent width of 5 \AA$\,$ may prove to be an
essentially arbitrary one \cite{ghi11_bllacs_fsrq}. As discussed in
\cite{geo98_thesis}, a line-less blazar may arise as a product of
beamed emission out-shining the spectral lines, which would then
emerge if the source were misaligned. A further complication arises
when considering the spectral types of radio galaxies.  High
excitation radio galaxies (HERG) have prominent emission lines and
presumably standard accretion disks accreting near the Eddington
limit, while low excitation radio galaxies (LERG) have few or no lines
and are likely radiatively inefficient. While most FR II are HERG and
most FR I are LERG, there are known exceptions in both cases,
including the nearby HERG FR I Cen A \citep{eva04_cena}, other
possible `FR I quasars' \citep{blu01,hey07_fri_qso,chi09_fri_cosmos},
and an increasing population of LERG FR II at higher redshifts (up to
z $\sim$ 0.5, \citealt{gar10}).

\begin{figure}
  \includegraphics[width=0.99\linewidth]{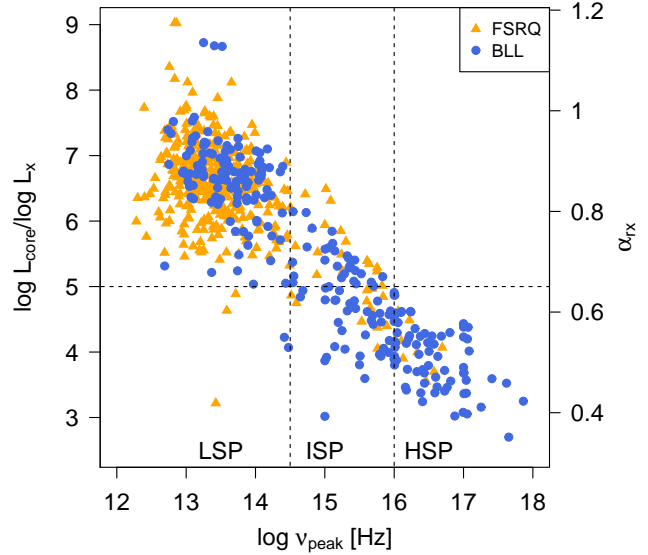}

  \caption{The ratio log
    ($\mathrm{L}_{\mathrm{core}}$/$\mathrm{L}_\mathrm{X}$) versus
    synchrotron peak frequency for all 653 sources with well-measured
    synchrotron peaks (equivalent $\alpha_{\mathrm{rx}}$ is shown on
    the right scale, $\alpha_{\mathrm{rx}}$ = log
    ($\mathrm{L}_\mathrm{core}/\mathrm{L}_\mathrm{X})$/7.68). This
    ratio is often used to estimate the location of the synchrotron
    peak frequency when detailed SED information is
    lacking. \cite{lan08} use log
    ($\mathrm{L}_{\mathrm{core}}$/$\mathrm{L}_\mathrm{X}) <$ 5 to
    indicate high synchrotron peak, however as shown, this ratio may
    only give a reliable upper limit to the frequency, and low log
    ($\mathrm{L}_{\mathrm{core}}$/$\mathrm{L}_\mathrm{X}$) values do
    not correspond exclusively to high synchrotron peaks. }
  
\label{fig:lclx}
\end{figure}

An explanation is given by \cite{gar10} for the presence of these
mixed-type AGN based on black hole spin $-$ namely, that central black
holes with high retrograde spins and efficient accretion will produce
powerful HERG FR IIs, which slowly spin down to prograde-spinning, low
efficiency FR Is.  However, if the transition between spin state
occurs significantly before or after the transition in accretion
efficiency, mixed-state objects will occur.  Interestingly, the
spin-based unification scenario may explain the lack of an obvious
continuum of blazar types, as in this scenario the transition objects
(near zero spin) have much lower jet powers than rapidly spinning FR I
or FR IIs, thus in our typical luminosity-dependent plots there will
be a `missing population' of transition objects, either showing up at
much lower powers or missing due to selection effects.

%-------------------------------------------------------------------------------
\begin{deluxetable*}{p{1.2in}p{2.6in}p{2.6in}}
  \tablecaption{Overview of Strong and Weak Jet Classes\label{table:ov}}  
  \tablehead{
    & Strong & Weak \\
  }
  \startdata
  Definition & More powerful, single-component jet (single characteristic $\Gamma$), remains relativistic on large scales & Lower-power jet with faster deceleration, multiple emitting regions (multiple $\Gamma$) \\[0.1in]
  Optical Type& mostly FSRQ & mostly BL Lacs\\[0.1in]
  Synchrotron Peak& All high-power LSP, some low-power LSP & All ISP, HSP, some low-power LSP\\[0.1in]
  Typical $\mathrm{L}_{300}$ & $10^{40}-10^{45}$ ergs s$^{-1}$ & $10^{38}-10^{42}$ ergs s$^{-1}$ \\[0.1in]
  Typical $\mathrm{L}_{\mathrm{kin}}$ & $10^{43.5}-10^{46}$ ergs s$^{-1}$ & $10^{42}-10^{44.5}$ ergs s$^{-1}$ \\[0.1in]
  Radio Morphology & FR II; Collimated jets, hot spots, edge-brightened & FR I; Less collimated, edge-darkened, no hot spots \\[0.1in]
  Accretion Type & Efficient, likely standard disk & Inefficient, highly sub-Eddington\\
  \enddata
\end{deluxetable*}
% -----------------

Ultimately, it is not clear what role the ionizing continuum or line
emitting regions has in influencing the jet type. Optical spectral type
is therefore not a part of our suggested strong/weak classification,
which relies instead on the source morphology, core-dominance, and
location in the $\nu_{\mathrm{peak}}-\mathrm{L}_{\mathrm{peak}}$
plane. Thus, many of the low-peaking sources with high core dominance
(\emph{i.e.}, strong jets) are BL Lacs, including BL Lac itself and
all but one of the nine hot-spot BL Lacs found in the MOJAVE sample
\citep{kha10}. Similarly, several quasars appear at higher peak
frequencies (\emph{i.e.}, as weak jets) in the UEX sample.

Lastly, we comment on the possible presence of high-frequency-peaking
FSRQ (HFSRQ) as claimed by several authors
\citep{lan06,pad07_dxrbs_iii,lan08_quasars}. While the presence of
quasars with low bolometric luminosities and high peaks (essentially
having SEDs typical of HSP BL lacs) are consistent with expectations
based on the updated theoretical sequence
\citep{ghi08_new_perspective}, many of the objects discussed do have
higher jet kinetic powers typical of FR IIs. However, using low core
radio to X-ray flux (log $\mathrm{L}_{\mathrm{core}}$/log
$\mathrm{L}_\mathrm{X}$ $<$ 5) or high optical to X-ray ratio
($\alpha_{ox}$) is not a reliable way to find sources with high
$\nu_{\mathrm{peak}}$.  To illustrate this, in Figure \ref{fig:lclx}
we plot log ($\mathrm{L}_{\mathrm{core}}$/$\mathrm{L}_\mathrm{X}$),
measured at 5 GHz and 1 keV, against the known peak from our SED fits
(based on much more extensive data than a two-point spectral index)
for the 653 sources in our TEX+UEX sample.  While a few quasars exist
in the HSP range $>10^{16}$ Hz, these sources appear otherwise very
similar to HSP BL Lacs in their SEDs.  Further, many sources below the
log ($\mathrm{L}_{\mathrm{core}}$/$\mathrm{L}_\mathrm{X}$) $<$ 5 limit
clearly have low peak values, often with a hard X-ray spectrum
indicating a non-synchrotron origin. So far no unambiguous evidence of
any blazar with a high peak frequency \emph{and high luminosity}
(above the typical HSP zone) has been found
\citep[see][]{mar08,lan08_quasars}.

\subsubsection{Intermediate and Low-peaking Blazars}

According to the proposed two-population scenario, all ISP and HSP
sources belong to the weak jet parent sample along with FR I radio
galaxies. As blazars, they move from a population of HSP to ISP and
finally LSP sources as they are misaligned.  Contrary to the
interpretation of a continuous sequence \citep{fos98}, then, ISP
blazars should have lower core dominance, lower overall luminosities,
lower Doppler factors, and yet similar $\mathrm{L}_\mathrm{kin}$ when
compared with HSP sources. However, it is difficult to rule out a
missing population of intermediate sources which may fill in the gap
between low and high peaks in Figure \ref{fig:cdvp} or in the center
of Figure \ref{fig:envelope}. These less populated regions are not due
to the flux-limited selections of the original master sample (see
below). If these sources are in the master sample, the SED sampling
would have to be much lower, and for such relatively bright objects,
this seems unlikely.

Weak jet sources at even greater angles to the line of sight should
appear as LSP blazars, though they may not be recognizable as such
(but rather as radio galaxies). Regardless, these should be
distinguishable from strong jet LSP sources based on core dominance,
morphology, extended radio power, and polarization signatures.
However without detailed VLBI and VLA studies to measure these
attributes, confusion among the two LSP populations is likely to
occur. From our hypothesis that $\mathrm{R}_\mathrm{CE}$ must decrease
with peak frequency for weak-jet sources, we expect most of the
low-power LSP blazars shown in Figure \ref{fig:cdvp} to belong to the
parent sample of strong jet sources, along with FR II radio
galaxies. This connection is suggested by the equal or higher core
dominance of most of these sources compared with the HSP blazars in
the same $\mathrm{L}_\mathrm{kin}$ range. It is also supported by the
detailed observations we do have from VLBI studies of some of our
objects, such as the nine LSP BL Lacs in the MOJAVE sample with
hotspots similar to those typically seen in high-powered quasars and
associated with FR II morphology \citep{kha10}. The MOJAVE blazars
(almost entirely LSP sources) were also found to have an intrinsic,
unbeamed luminosity function consistent with that of FR II radio
galaxies, whether BL Lacs were included or not \citep{car08}.

We summarize the strong and weak jet characteristics as suggested by
the current work in Table \ref{table:ov}.

\subsection{Examining the Sample Bias}
\label{section:biases}
Our effort is concerned with those sources which can be placed on the
$\nu_{\mathrm{peak}}-\mathrm{L}_{\mathrm{peak}}$ plane with
confidence, which is primarily a function of SED sampling. It is
important to examine both the selection of the entire sample and the
nature of the discarded objects for evidence of any bias which might
produce the current findings.

\begin{figure}
    \includegraphics[width=0.99\linewidth]{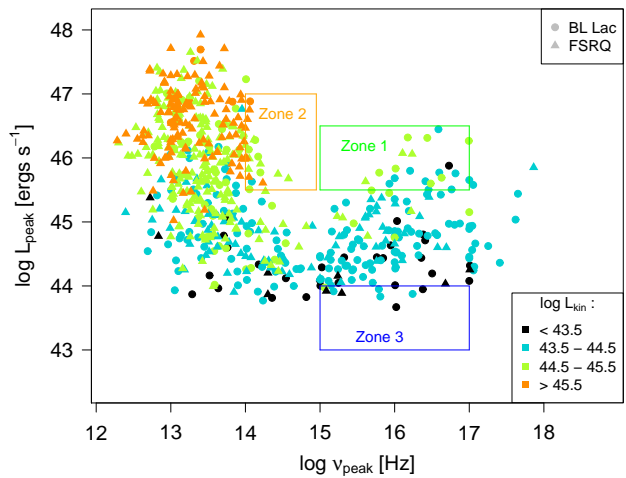}
    \caption{The blazar envelope for the UEX sample of 437 sources with
      only upper limits on $\mathrm{L}_\mathrm{kin}$. Upper limits
      were found by extrapolating from the lowest frequency
      observation of core flux using a spectral index $\alpha$ =
      1.2. The same color scale is used as in Figure
      \ref{fig:envelope}. We clearly see a similar `L' shape as is
      suggested in the figure for TEX sources.  Also shown are the
      three zones discussed in \S \ref{section:biases}.}
\label{fig:upperlim}
\end{figure}

One possible selection bias is a `windowing' effect due to the low
frequency radio measurements required to isolate the core and extended
components. On one side, very core-dominated sources are likely to be
missed when using only the spectral decomposition method, as these
might not exhibit a steep spectrum except below very low frequencies
(10 $-$ 100 MHz) which are not frequently observed (and thus fail the
two-component test). On the other side, a very de-beamed source may
not qualify as a flat-spectrum source if the $\nu_{\mathrm{cross}}$ is
much above typical sample definition frequencies of a few GHz. The
combined effect could in principle lead to the selection of sources
for which the core and extended fluxes are by constraint rather
similar.  The inclusion of VLA map data, which allows us to find
extended flux densities even in very core-dominated sources, should
mitigate the first issue. As to the second, as we show in Figure
\ref{fig:xcross}, most blazars are core dominated above the crucial
window of a few GHz, and we detect a few above this limit.  Between
the blazars and radio galaxies in this figure there is no large gap
indicating a missing population, and we do not find any reasons to
suspect that the windowing effect is a problem for the results of this
paper.

\subsubsection{Upper Limits}
The over 400 sources in the UEX sample are shown in Figure
\ref{fig:upperlim}, with the color corresponding to the upper limit of
the jet kinetic power. The overall shape of the envelope appears
qualitatively similar to Figure \ref{fig:envelope}.

We tested areas of the
$\nu_{\mathrm{peak}}-\mathrm{L}_{\mathrm{peak}}$ plane using
simulations to check how the sensitivity to sources in the plane is
affected by our overall combined sample selection (\emph{i.e.,} the
combined selection of the samples in Table \ref{table:sources}). We
moved $\sim$100 template SEDs drawn from our database to various peak
locations in each zone shown in Figure \ref{fig:upperlim} and
simulated cuts using the flux limits of our individual samples.  In
zones 1 and 2, we did not find any difficulty detecting the simulated
sources out to a reasonable redshift (z $\sim$2). In zone 3, however,
our sample would likely only include sources out to z $\sim 0.2$,
which is not much less than the typical redshift of the HSP sources
just above this region (z $\sim 0.25$).  We conclude that the
particular lack of ISP sources with $\mathrm{L}_{\mathrm{peak}}$
intermediate between LSP and HSP sources is not due to the
\emph{initial} selection of our sample. For at least zones 1 and 2,
which would contain very bright sources, it is unlikely that a large
number of sources would exist there, but lack the SED sampling
necessary to make the more subjective cuts leading to our final
sample.

\subsubsection{Remaining Sample}
\label{sec:discards}
We now examine the sources which we were unable to include due to
inadequate SED sampling (typically fewer than $\sim$ 10 SED data
points). Examination of the colors (radio-optical and optical-X-ray
indices) for these sources indicates that they have values consistent
with the well-characterized sample of $\sim$ 600 TEX+UEX
sources. Unsurprisingly, those with X-ray detections mostly populated
the `HSP zone' of the $\alpha_{ro}-\alpha_{ox}$ plane, indicating that
a large population of the excluded sources are likely higher redshift
HSPs and ISPs. 

The optical and X-ray luminosity distributions for the excluded sample
as a whole are indistinguishable (by KS test) from the TEX+UEX sample.
In all three bands (and in particular the radio), those sources
detected in the X-rays tended to have much lower luminosities, and are
therefore unlikely to populate the so-called `forbidden zone' of high
synchrotron peak luminosity and frequency.  Those not detected in
X-rays had a distribution of $\alpha_{\mathrm{ro}}$ typical of LSP
sources, but actually tended to have higher luminosities than our
TEX+UEX sample.  However, non-detection in X-rays and higher redshifts
in general probably explains their absence in the TEX+UEX
sample. Overall, none of the sources excluded from the TEX+UEX
sub-sample appear to be substantially different from those in it;
however, only additional multi-frequency observations of these sources
can confirm this.

\section{SUMMARY AND CONCLUSIONS}
\label{conclusions}
In this work we have analyzed the multi-frequency spectrum of several
hundred blazars in order to accurately measure the synchrotron peak
frequency and luminosity, and performed spectral analysis of the
low-frequency emission in order to estimate the intrinsic jet power by
a scaling of the 300 MHz extended radio power. The connection of
blazars and radio galaxies in the same plane amounts to an extension
of the original blazar sequence to a blazar envelope, in which highly
aligned blazars form an upper limit in the
$\nu_{\mathrm{peak}}-\mathrm{L}_{\mathrm{peak}}$ plane, while radio
galaxies populate the lower left. We find that the jet power does not
absolutely determine the $\nu_\mathrm{peak}$ of a maximally-beamed
source. Instead, there seems to a break-down at low kinetic jet power,
such that these jets are capable of having either low or high
synchrotron peaks when aligned with the line of sight.

Our results suggest a shift away from the continuous blazar sequence
interpretation, but confirmation waits for ever more data.
%We anticipate a follow-up publication after updates from new missions like \emph{Planck} and \emph{WISE} which are expected to bring many more sources into our trusted sample. 
We summarize the most important points here:

\begin{itemize}
\item Among our sample of over 600 blazars with well-sampled blazar
  SEDs, none with high peak frequencies and high luminosities are
  found. This void may be explained by the original arguments advanced
  by \cite{ghi98}, that higher external radiation fields in high-power
  sources lead to greater cooling and thus lower break/peak energies.

\item However, it is not clear that there is a continuum of SED types
  with jet power. The source power alone does not appear to determine
  the placement of a source along the aligned sequence.  Instead, we
  suggest that radio-loud AGN can be divided into two populations:
  low-synchrotron-peaking sources without significant velocity
  gradients or deceleration (strong jets) and high-synchrotron-peaking
  sources which do have such structure (weak jets).  The attributes of
  these samples are summarized in Table \ref{table:ov}.
  
\item It appears that below a certain jet power, sources can either
  have strong or weak jets, suggesting another parameter besides
  kinetic power is responsible for the jet structure. As in
  \cite{ghi08_new_perspective}, we suggest that this parameter is the
  accretion rate $\dot{m}$ in Eddington units with the two families
  separated at $\dot{m}=\dot{m}_{cr}\sim 3\times 10^{-3}-10^{-2}$,
  with sources having $\dot{m}<\dot{m}_{cr}$ exhibiting radiatively
  inefficient accretion \citep{narayan97}.

\item A key prediction of the two-population scenario is that ISP
  sources primarily consist of misaligned HSP objects, as suggested by
  their slightly lower $\mathrm{L}_{\mathrm{peak}}$ values and similar
  $\mathrm{L}_\mathrm{kin}$ to HSPs. Observations of ISP sources with
  synchrotron peak values and core dominance values between those
  typical of the LSP and HSP sources (currently missing) would support
  a continuous sequence interpretation instead.
  
\item Observations of highly core-dominated, low-peaking BL Lacs (such
  as BL Lac itself), including several BL Lacs exhibiting FR II-like
  morphology, leads us to speculate that these sources belong to the
  strong-jet parent population, along with FR II radio
  galaxies. Likewise, some ISP and HSP sources with quasar-like
  spectra in our sample belong to the weak population. Thus the
  spectral types associated with strong and weak jets are mixed.
%\item While FR II radio galaxies have higher peak luminosities than FR
%  I, all FR radio galaxies examined here had very similar peak
%  frequencies, typically $< 10^{14}$ Hz.
\item The low peak frequencies seen in our FR I radio galaxies
  supports the interpretation of a horizontal movement in the
  $\nu_{\mathrm{peak}}-\mathrm{L}_{\mathrm{peak}}$ plane for weak
  sources which supports a model with velocity profiles in the jet and
  leads to our two-population scheme.  Observations of high
  synchrotron peaks in radio galaxies would be very difficult to
  explain under this scenario.
%\item Inverse Compton emission for our sources increases with both
%  $\mathrm{L}_\mathrm{kin}$ and the degree of beaming (measured
%  through $\mathrm{R}_\mathrm{CE}$). We also find that
%  $\mathrm{L}_\gamma$ for ISP blazars is similar to or perhaps lower
%  than $\mathrm{L}_\gamma$ of HSP sources, which may be explained as
%  another consequence of greater misalignment of ISP blazars.
\end{itemize}

\acknowledgments 

The authors wish to thank the anonymous referee for helpful comments,
and Katherine Blundell for a discussion which lead to the improvement
of this manuscript. This research has made use of the SIMBAD database,
operated at CDS, Strasbourg, France
(http://simbad.u-strasbg.fr/simbad/), and the NASA/IPAC Extragalactic
Database (NED) which is operated by the Jet Propulsion Laboratory,
California Institute of Technology, under contract with the National
Aeronautics and Space Administration
(http://nedwww.ipac.caltech.edu/). We utilized package \textit{np}
\citep{Rnp} in the \textit{R} language and environment for statistical
computing \citep{Rref}. GF acknowledges support from NASA grants
NNG05GJ10G, NNX06AE92G, and NNX09AR04G, as well as SAO grants
GO3-4147X and G05-6115X. MG acknowledges support from the NASA ATFP
grant NNX08AG77G and NASA FERMI grant NNH08ZDA001N. The MOJAVE project
is supported under National Science Foundation grant 0807860-AST.

%\clearpage

\appendix

\section{A. Phenomenological Blazar SED model}
\label{app:SEDfit}
The model was adapted from \cite{fos97} and used to fit the average
blazar SED with seven parameters, six of which were free to vary and
which were optimized using an SA routine.

\noindent
Parameters:
\begin{equation}
  \begin{array}{l l l l}
    &\qquad \qquad \qquad \qquad \qquad \qquad \qquad \qquad \qquad &&\\
    % \nonumber
    L_{R} = &32 - 46\nonumber & & \\
    \tau \equiv & 1000 & & \\
     x_{\mathrm{j}} = & 1 - 4 \nonumber &&\\
    \kappa_{\mathrm{rx}} = & 3 - 9 \nonumber&& \\
    \alpha_x = & -0.4 - 0.8 \nonumber && \\
     x_{\mathrm{peak}} = & 10 - 20 \nonumber&&\\
    \beta = & 0.6 - 2.0 \nonumber&&\\
  \end{array}
  %\right.
\end{equation}

\noindent
To build the SED, the following equations were combined, where
$\vec{x}$ = log($\vec{\nu}$). Vector quantities are noted with arrows for
clarity, and luminosities are in log.

\begin{equation}
\sigma = \sqrt{\frac{2\, x_\mathrm{j}}{\beta}}
\end{equation}

\begin{equation}
L_\mathrm{peak} = \beta\,( x_\mathrm{peak} -  x_\mathrm{j} - 9.698) + L_\mathrm{R}\,+\, \left(\frac{1}{2}\,\beta\, x_\mathrm{j}\right)
\end{equation}

\begin{equation}
  \vec{c}_\mathrm{damp} = 
  \left\{
  \begin{array}{l l}
    1 & \mbox{when $\vec{ x} <  x_\mathrm{peak}$}\\
    2 - 10^{\frac{\vec{ x} -  x_\mathrm{peak}}{\tau}} \quad \quad & \mbox{else}\\
  \end{array}
  \right.
\end{equation}

\begin{equation}  
  \vec{L}_\mathrm{synch} =  
  \left\{
    \begin{array}{l l}
      \beta\left(\vec{ x} - 9.698\right) + L_\mathrm{R} & \mbox{when $\vec{ x}\, <\, \left( x_\mathrm{peak} -  x_\mathrm{j}\right)$}\\ 
      \quad \quad&\quad \quad\\ 
      \left[L_\mathrm{peak} - \left(\frac{\vec{ x} -  x_\mathrm{peak}}{\sigma}\right)^2\right]\, \vec{c}_\mathrm{damp}\quad \quad& \mbox{else}\\
    \end{array}
    \right. \quad
\end{equation}

\begin{equation}
  \vec{L}_\mathrm{IC} = L_R - \kappa_\mathrm{rx} + 7.685 + \left(1 - \alpha_x\right)\left(\vec{x} - 17.383\right)
\end{equation}

\vspace{10pt}
\noindent
Finally, we have the complete SED:
\vspace{10pt}
\begin{equation}
\vec{L} = \mbox{log}\left[10^{\vec{L}_\mathrm{synch}} + 10^{\vec{L}_\mathrm{IC}}\right]
\end{equation}

\section{B. Radio Core Dominance}
\label{app:deriv}
Equation \ref{eq:lpcd} can be derived starting from the definition of
log core dominance ratio $\mathrm{R}_\mathrm{CE}$:

\begin{equation}
\mathrm{R}_\mathrm{CE} =
\mbox{log }\left(\frac{\mathrm{L}_\mathrm{R}}{\mathrm{L}_\mathrm{ext}}\right)
\end{equation}

\noindent
where $\mathrm{L}_\mathrm{ext}$ and $\mathrm{L}_\mathrm{R}$ are the
luminosities of the extended and core components taken at the same
frequency, and the beaming equations for the luminosity for the radio
and synchrotron peak frequency are simply:

\begin{equation}
\mathrm{L}_\mathrm{R} = \mathrm{L}_\mathrm{R}^\prime\delta_\mathrm{R}^{p_R}
\end{equation}
\begin{equation}
\mathrm{L}_\mathrm{peak} = \mathrm{L}_\mathrm{peak}^\prime\delta_\mathrm{peak}^{p_\mathrm{peak}}
\label{eq:b2}
\end{equation}

\noindent
where $\delta_\mathrm{R}$ and $\delta_\mathrm{peak}$ are the Doppler
factors for the jet regions dominating the emission at those
frequencies, and the exponents take values p = 2+$\alpha$ or 3+$\alpha$, depending on the jet model, and with $\alpha$ the spectral
indices in the band of interest.  By taking advantage of the
relationship $\mathrm{L}_\mathrm{kin}$ = $\kappa\mathrm{L}_\mathrm{ext}^\beta$ (see \S \ref{sec:CD}) we can
rewrite it as:

\begin{equation}
\mathrm{R}_\mathrm{CE} =
\mbox{log }\left(\frac{\mathrm{L}_\mathrm{R}^\prime\delta_\mathrm{R}^{p_R}}{\left(\frac{\mathrm{L}_\mathrm{kin}}{\epsilon}\right)^{1/\beta}}\right)
\end{equation}
\noindent
If we assume that $\mathrm{L}_\mathrm{R}^\prime$ =
$\epsilon \mathrm{L}_\mathrm{kin}$ (see \S \ref{sec:CD}), and that the
bulk Lorentz factors $\Gamma$ for the radio and the peak emission are
the same, such that $\delta_\mathrm{R}$ = $\delta_\mathrm{peak}$, using \ref{eq:b2}, we obtain:

\begin{equation}
  \mathrm{R}_\mathrm{CE} =
  \mbox{log }\left(\epsilon\mathrm{L}_\mathrm{kin}\left(
  \frac{\mathrm{L}_\mathrm{peak}}
       {\mathrm{L}_\mathrm{peak}^\prime}\right)^{\frac{p_R}{p_\mathrm{peak}}}\right)
       - \mbox{log }\left(\frac
       {\mathrm{L}_\mathrm{kin}}
       {\kappa}\right)^{1/\beta}
\end{equation}
\noindent
Letting $p_R$ = 2 + $\alpha_R$
and $p_\mathrm{peak}$ = 2 + $\alpha_\mathrm{peak}$ = 3, we have equation \ref{eq:lpcd}:

\begin{equation}
\mathrm{R}_\mathrm{CE} = \left(\frac{2+\alpha_R}{3}\right)\mbox{log }\mathrm{L}_\mathrm{peak} + \left(1-\frac{1}{\beta}\right)\mbox{log }\mathrm{L}_\mathrm{kin} + c_2
\end{equation}

\begin{equation}
  c_2 = \mbox{log}(\epsilon
  \kappa^{1/\beta}) - \left(\frac{2+\alpha_R}{3}\right)\mbox{log }\mathrm{L}_\mathrm{peak}^\prime
\end{equation}

\bibliographystyle{apj}
\bibliography{apj-jour,bib_be_letter,extra}

\end{document}